\documentclass[]{jfm}

\usepackage[dvipsnames]{xcolor}

\usepackage{graphicx}
\usepackage{newtxtext}
\usepackage{newtxmath}
\usepackage{natbib}
\usepackage{hyperref}
\hypersetup{
    colorlinks = true,
    urlcolor   = blue,
    citecolor  = black,
}

\newcommand{\RomanNumeralCaps}[1]
\linenumbers

\usepackage{graphicx}
\usepackage{epstopdf, epsfig}
\usepackage{amsmath}
\usepackage[utf8]{inputenc} 

\usepackage{subfigure}% in preamble

\usepackage{pgfplots}
\usepackage{tikz}
\usetikzlibrary{backgrounds,pgfplots.groupplots,external}

\title{Super-harmonically resonant swirling waves in longitudinally forced circular cylinders} 

\author{ Alice Marcotte, 
         François Gallaire 
         \corresp{\email{francois.gallaire@epfl.ch}}
    \and Alessandro Bongarzone
       }

\affiliation{\aff{1}Laboratory of Fluid Mechanics and Instabilities, École Polytechnique Fédérale de Lausanne, Lausanne, CH-1015, Switzerland}

\begin{document}

\maketitle
%\tableofcontents

\begin{abstract}

Resonant sloshing in circular cylinders was studied by \cite{faltinsen2016resonant}, whose theory was used to describe steady-state resonant waves due to time-harmonic container's elliptic orbits. In the limit of longitudinal container motions, a symmetry-breaking of the planar wave solution occurs, with clockwise and anti-clockwise swirling equally likely. In addition to this primary harmonic dynamics, previous experiments have unveiled that diverse super-harmonic dynamics are observable far from primary resonances. Among these, the so-called double-crest (DC) dynamics, first observed by \cite{reclari2014surface} for rotary sloshing, is particularly relevant, as its manifestation is the most favored by the spatial structure of the external driving. Following \cite{bongarzone2022amplitude}, in this work we develop a weakly nonlinear (WNL) analysis to describe the system response to super-harmonic longitudinal forcing. The resulting system of amplitude equations predicts that a planar wave symmetry-breaking via stable swirling may also occur under super-harmonic excitation. This finding is confirmed by our experimental observations, which identify three possible super-harmonic regimes, i.e. (i) stable planar DC waves, (ii) irregular motion and (iii) stable swirling DC waves, whose corresponding stability boundaries in the forcing frequency-amplitude plane quantitatively match the present theoretical estimates.
\end{abstract}


%%%%%%%%%%%%%%%%%%%%%%%%%%%%%%%%%%%%%%%%%%%%%%%%%%%%%%%%
%%%%%%%%%%%%%%%%%%%%%%%%%%%%%%%%%%%%%%%%%%%%%%%%%%%%%%%%
%%%%%%%%%%%%%%%%%%%%%%%%%%%%%%%%%%%%%%%%%%%%%%%%%%%%%%%%
%%%%%%%%%%%%%%%%%%%%%%%%%%%%%%%%%%%%%%%%%%%%%%%%%%%%%%%%

\section{Introduction}\label{sec:Sec1}

Liquid sloshing related problems remain nowadays of great concern to many engineering fields. Depending on the type of external forcing and container shape, the free liquid surface can experience different types of dynamics, whose nature has a major importance in the design of, e.g., airplanes, rockets, spacecraft as well as road and ship tankers, since the sloshing motion may have a strong influence on their dynamic stability \citep{ibrahim2009liquid,faltinsen2005liquid}. The case of resonant sloshing in upright circular cylinders represents one of the archetypal sloshing systems and it has indeed been extensively studied theoretically, experimentally and numerically.\\
\indent In their work, \cite{faltinsen2016resonant} thoroughly examine harmonically resonant sloshing dynamics in upright annular (circular) reservoirs. By applying the Narimanov–Moiseev multimodal sloshing theory \citep{narimanov1957movement,moiseev1958theory,dodge1965liquid,faltinsen1974nonlinear,narimanov1977,lukovsky1990,lukovsky2011combining,lukovsky2015multimodal,lukovsky2015mathematical,takahara2012frequency}, capable of accurately describing the nonlinear wave dynamics near primary harmonic resonances and in absence of secondary resonances \citep{faltinsen2005resonant,faltinsen2016resonant,raynovskyy2018steady,raynovskyy2020sloshing}, \textcolor{black}{i.e. for a non-dimensional fluid depth $H\gtrsim1.05$}, they derived the response curves for planar elliptic-type tank excitation. In the two limit cases, system responses to longitudinal and rotary tank motions were retrieved.\\
\indent Rotary sloshing is widely used in biological and chemical industrial applications such as small and large scale bioreactors for bacterial and cellular cultures \citep{mcdaniel1969effect,wurm2004production}, where the liquid motion prevents the sedimentation of suspended cells in the liquid medium and allows for a homogenized concentration of dissolved oxygen and nutrients. For these reasons, a strong interest in the gas exchange and mixing processes taking place in these devices has emerged over the last decades \citep{buchs2000power1,buchs2000power2,buchs2001introduction,maier2004advances,muller2005orbital,micheletti2006fluid,zhang2009efficient,tissot2010determination,tan2011measurement,tissot2011efficient,klockner2012advances}.\\
\indent \cite{reclari2013hydrodynamics} and \cite{reclari2014surface}, among others (see also  \cite{hutton1964fluid,bouvard2017mean,moisy2018counter,horstmann2020linear,horstmann2021formation}), experimentally characterized in great detail the hydrodynamics of orbitally shaken circular cylinders, which represent the typical shape of lab-scale bioreactors. In addition to the primary harmonic system response via single-crest (SC) wave dynamics, different multiple-crest wave patterns were observed. Among these, the super-harmonic double-crest (DC) wave dynamics, as labeled by \cite{reclari2014surface}, is particularly relevant, as it appears to be the most stable and the one which displays the largest nonlinear amplitude response, that may eventually lead to wave breaking occurring far from harmonic resonances and even at moderately low forcing amplitudes. Its manifestation is indeed naturally favored by the spatial structure, i.e. by the temporal and azimuthal periodicities, of the external driving and, therefore, its understanding and prediction can be important for practical application \textcolor{black}{as in the design of bioreactors.}\\ 
\indent The analysis outlined in \cite{bongarzone2022amplitude} was precisely dedicated to the development of a inviscid weakly nonlinear analysis, which was seen to successfully capture nonlinear effects for this subtle additive and multiplicative resonance governing the super-harmonic double-crest swirling and which well matched the experimental findings of \cite{reclari2013hydrodynamics} and \cite{reclari2014surface}.\\ 
\indent Nonetheless, the applicability of the aforementioned analysis is limited to rotary sloshing, whereas the emergence of super-harmonic DC dynamics is in principle expected for any elliptic-type container excitation and, therefore, for longitudinal forcing as well.\\
\indent The latter forcing condition has been analytically and experimentally studied for decades \citep{hutton1963inv,abramson1966dynamic,chu1968subharmonic} and it is of interest from the perspective of hydrodynamic instabilities due to the occurrence of hysteretic symmetry-breaking conditions \citep{miles1984internally,miles1984resonantly}. With regards to \textcolor{black}{circular cylindrical containers}, particularly relevant are the experimental studies by \cite{abramson1966some}, \cite{royon2007liquid} and \cite{hopfinger2009liquid}, who detected the stability bounds between harmonic planar, swirling and irregular waves and whose estimates were later used by \cite{faltinsen2016resonant} to validate their theoretical analysis. However, these works were mostly focused on the investigation of system responses in the neighborhood of harmonic resonances, whereas, with the exception of \cite{reclari2014surface} and \cite{bongarzone2022amplitude} in the context of rotary sloshing, the literature seems to lack of comprehensive experimental and theoretical studies dealing with the most relevant secondary super-harmonic resonances (by super-harmonic, we mean here a wave of a certain frequency $\omega$ emerging from an excitation at $\Omega = \omega/2$, with $\Omega$ the driving angular frequency), i.e. far from primary ones, under longitudinal or, more generally, elliptical container excitation.\\
\indent In this work we take a first step in this direction by extending to longitudinal planar forcing the analysis formalized by \cite{bongarzone2022amplitude} for circular container motions. In the spirit of the multiple timescale method, we develop a weakly nonlinear (WNL) model leading to a system of two amplitude equations, which, via thorough comparison with dedicated lab-scale experiments, is proven capable of describing satisfactorily the steady-state system response to super-harmonic longitudinal forcing and, particularly, of detecting the various possible dynamical regimes.\\
\indent The manuscript is organized as follows. The flow configuration and governing equations are given in \S\ref{sec:Sec2}. In \S\ref{sec:Sec3} we briefly introduce the classical linear potential model together with a short description of the numerical method employed in this work. By analogy with \cite{bongarzone2022amplitude}, in \S\ref{sec:Sec4}, we first tackle the simpler case of harmonic single-crest (SC) wave. The WNL system of amplitude equations governing the double-crest (DC) wave dynamics under super-harmonic longitudinal forcing, which represents the core of this study, is then formalized in \S\ref{sec:Sec5}. The experimental apparatus, procedure and findings are described in \S\ref{sec:Sec6}, where a thorough quantitative comparison with the present theoretical estimates is carried out. Final comments and conclusions are outlined in \S\ref{sec:Sec7}. Lastly, Appendix~\ref{sec:AppC} complements the theoretical model by briefly showing how a straightforward extension of the present analysis to generic container's elliptic orbits can be readily obtained without any further calculation, hence paving the way for further analyses and experimental investigations.

\section{Flow configuration and governing equations}\label{sec:Sec2}

%2.1614-2.0943951

% CONTAINER SKETCH 
\begin{figure}
\centering
\includegraphics[scale=0.4]{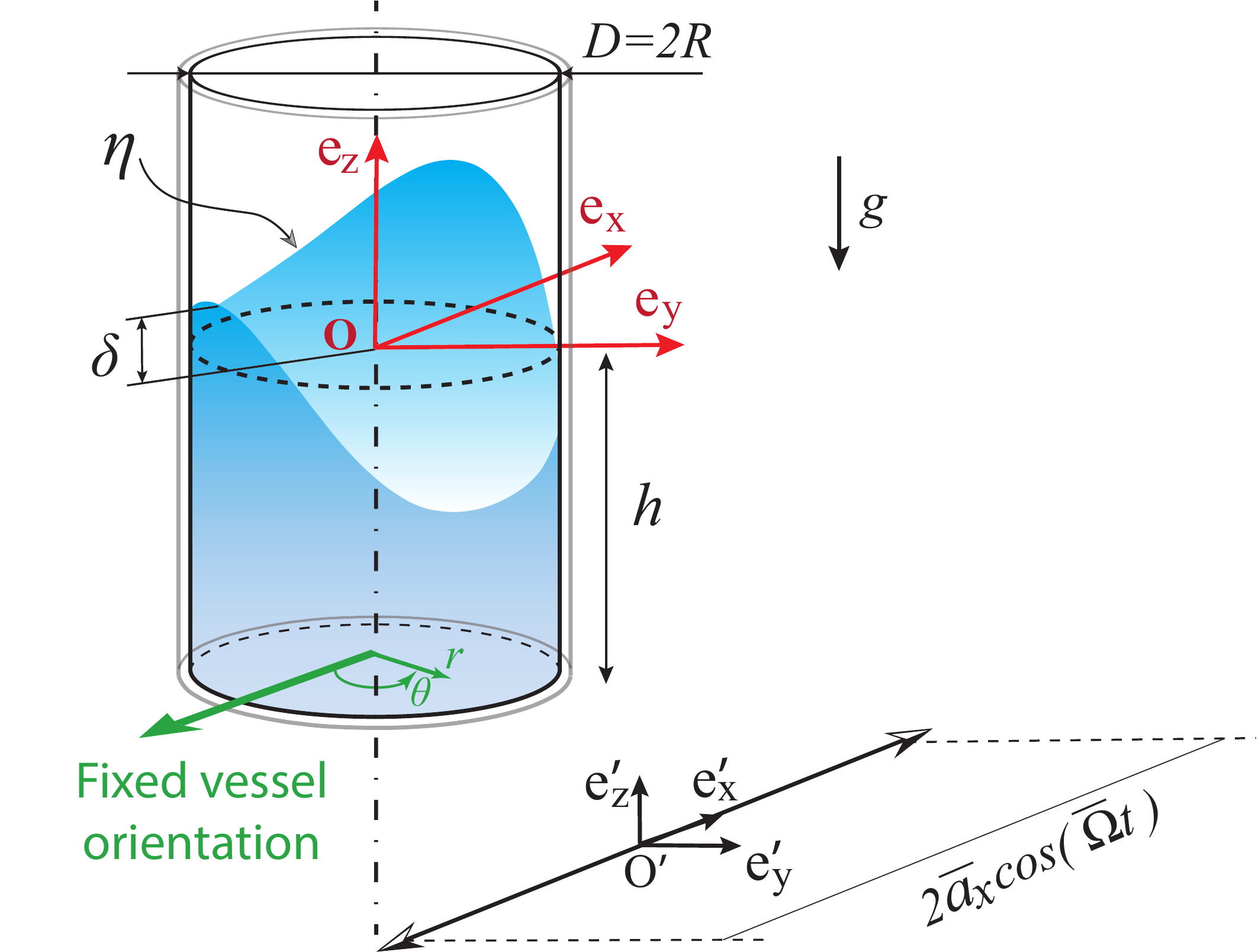}% Here is how to import EPS art
\caption{Sketch of a cylindrical container of diameter $D=2R$ and filled to a depth $h$. The gravity acceleration is denoted by $g$. $O'\mathbf{e}'_x\mathbf{e}'_y\mathbf{e}'_z$ is the Cartesian inertial reference frame, while $O\mathbf{e}_x\mathbf{e}_y\mathbf{e}_z$ is the Cartesian reference frame moving with the container. The origin of the moving cylindrical reference frame $\left(r,\theta,z\right)$ is placed at the container revolution axis and, specifically, at the unperturbed liquid height, $z=0$. The perturbed free surface and contact line elevation are denoted by $\eta$ and $\delta$, respectively. $\bar{a}_x$ is the amplitudes of the longitudinal periodic forcing of driving angular frequency \textcolor{black}{$\bar{\Omega}$}.}
\label{fig:Fig0} 
\end{figure}

We consider a cylindrical container of diameter $D=2R$ filled to a depth $h$ with a liquid of density $\rho$. The air--liquid surface tension is denoted by $\gamma$, whereas the gravity acceleration is denoted by $g$. $O'\mathbf{e}'_x\mathbf{e}'_y\mathbf{e}'_z$ is the Cartesian inertial reference frame, while $O\mathbf{e}_x\mathbf{e}_y\mathbf{e}_z$ is the Cartesian reference frame moving with the container. The origin of the moving cylindrical reference frame $\left(r,\theta,z\right)$ is placed at the container revolution axis and, specifically, at the unperturbed liquid height, $z=0$ (see figure~1). A longitudinal shaking in the horizontal plane, e.g. along the $x$-axis, can be represented by the following equations describing the motion velocity of the container axis intersection with the $z=0$ plane, parametrized in polar coordinates ($r$, $\theta$),
\begin{equation}
\label{eq:EqMotWall}
\dot{\mathbf{X}}_0 =
  \begin{cases}
    \, \, -\, \bar{a}_x\textcolor{black}{\bar{\Omega}}\sin{\left(\textcolor{black}{\bar{\Omega} }t\right)}\cos\theta\,\mathbf{e}_r\\
   \, \, \, \, \, \, \, \bar{a}_x\textcolor{black}{\bar{\Omega}}\sin{\left(\textcolor{black}{\bar{\Omega}} t\right)}\sin\theta\,\mathbf{e}_{\theta}
  \end{cases},
\end{equation}
with $\bar{a}_x$ the dimensional forcing amplitude and \textcolor{black}{$\bar{\Omega}$} the dimensional driving angular frequency. In the potential flow limit, the liquid motion within the moving container is governed by the Laplace equation, subjected to the homogeneous no-penetration condition at the solid lateral wall and bottom,
\begin{equation}
\label{eq:GovEq_Lap}
\Delta\Phi=0,\ \ \ \ \nabla\Phi\cdot\mathbf{n}=\mathbf{0},
\end{equation}
and by the dynamic and kinematic boundary conditions at the free surface $z=\eta\left(r,\theta\right)$ \citep{ibrahim2009liquid,faltinsen2005liquid},
\begin{subequations}
\begin{equation}
\label{eq:GovEq_Dyn}
\frac{\partial\Phi}{\partial t}+\frac{1}{2}\nabla\Phi\cdot\nabla\Phi+\eta-\frac{\kappa\left(\eta\right)}{Bo}=rf\cos{\left(\Omega t\right)}\cos{\theta},
\end{equation}
\begin{equation}
\label{eq:GovEq_Kin}
\frac{\partial \eta}{\partial t}+\nabla\Phi\cdot\nabla\eta-\frac{\partial\Phi}{\partial z}=0.
\end{equation}
\end{subequations}
\noindent which have been made non-dimensional by using the container's characteristic length $R$, the velocity $\sqrt{gR}$ and the time scale $\sqrt{R/g}$. In~\eqref{eq:GovEq_Dyn}, $\kappa\left(\eta\right)$ denotes the fully nonlinear curvature, while $Bo=\rho gR^2/\gamma$ is the Bond number. As soon as the Bond number is sufficiently large, i.e. $Bo\sim10^3$ \citep{bouvard2017mean}, surface tension effects are almost negligible (fully negligible for $Bo \gtrsim 10^4$, except in the neighborhood of the contact line \citep{faltinsen2016resonant}). In the following, we assume large Bond numbers and accordingly the curvature term in~\eqref{eq:GovEq_Dyn} is neglected. The non-dimensional driving acceleration along the $x$-axis reads $f=a_x\Omega^2$, with $a_x=\bar{a}_x/R$ and $\Omega=\textcolor{black}{\bar{\Omega}}/\sqrt{g/R}$. Lastly, the non-dimensional fluid depth is $H=h/R$.

\section{Linear potential model}\label{sec:Sec3}

Far from resonances and in the limit of small forcing amplitudes, the linear theory is expected to provide a good approximation of the harmonic system response. Let us consider small perturbations of the rest state,
\begin{equation}
\label{eq:straight_asymp_exp}
\mathbf{q}\left(r,\theta,z,t\right)=\left\{\Phi\left(r,\theta,z,t\right),\eta\left(r,\theta,t\right)\right\}^T=\epsilon \mathbf{q}'=\epsilon\left\{\Phi',\eta'\right\}^T+\text{O}\left(\epsilon^2\right),
\end{equation}
together with the assumption of small driving forcing amplitudes of order $\text{O}\left(\epsilon\right)$, i.e. $f=\epsilon F$, with $\epsilon$ a small parameter $\epsilon\ll 1$ and with the auxiliary variable $F$ of order $\text{O}\left(1\right)$. At order $\epsilon$, equations~\eqref{eq:GovEq_Lap}-\eqref{eq:GovEq_Kin} reduce to a forced linear system, whose matrix compact form reads,
\begin{equation}
\label{eq:LinMatForm}
\left(\partial_t\mathsfbi{B}-\mathsfbi{A}\right)\mathbf{q}'=\boldsymbol{\mathcal{F}}',
\end{equation}
with $\boldsymbol{\mathcal{F}}'=F\boldsymbol{\hat{\mathcal{F}}}\left(\frac{1}{2}e^{\text{i}\left(\Omega t-\theta\right)}+\frac{1}{2}e^{\text{i}\left(\Omega t+\theta\right)}\right)+c.c.$, $\boldsymbol{\hat{\mathcal{F}}}=\left\{0,\frac{r}{2}\right\}^T$ and
\begin{equation}
    \label{eq:LinMatForm_Expr}
    \mathsfbi{B} = 
 \begin{pmatrix}
  0 & 0 \\
  I_{\eta} & 0\\
 \end{pmatrix},\, \, \,
    \mathsfbi{A} = 
 \begin{pmatrix}
  \Delta & 0 \\
  0 & -I_{\eta}\\
 \end{pmatrix},
 \end{equation}
where $c.c.$ stands for complex conjugate and $I_{\eta}$ is the identity matrix associated with the interface $\eta$. Note that the kinematic condition does not explicitly appear in~\eqref{eq:LinMatForm_Expr}, but it is enforced as a boundary condition at the interface \citep{Viola2018a}. We then seek for a standing wave solution in the form 
\begin{equation}
\label{eq:GenEigProb0}
\mathbf{q}'\left(r,\theta,z,t\right)=F\hat{\mathbf{q}}\left(r,z\right)\left(\frac{1}{2}e^{\text{i}\left(\Omega t-\theta\right)}+\frac{1}{2}e^{\text{i}\left(\Omega t+\theta\right)}\right)+c.c.,
\end{equation}
where $\hat{\mathbf{q}}$ is straightforwardly computed by solving the system
\begin{equation}
\label{eq:GenEigProb1}
\left(\text{i}\Omega\mathsfbi{B}-\textcolor{black}{\mathsfbi{A}_{m=1}}\right)\hat{\mathbf{q}}=\hat{\boldsymbol{\mathcal{F}}},
\end{equation}
\noindent Note that, due to \textcolor{black}{the normal mode ansatz}~\eqref{eq:GenEigProb0}, the linear operator \textcolor{black}{$\mathsfbi{A}_m$} depends on the azimuthal wavenumber $m$, here $m=1$. Despite the fact that an exact analytical solution to equation~\eqref{eq:GenEigProb1} can be readily obtained via a Bessel-Fourier-series representation, in this work, as in \cite{bongarzone2022amplitude}, we opt for a numerical scheme based on a discretization technique, where linear operators $\mathsfbi{B}$ and \textcolor{black}{$\mathsfbi{A}_{m}$} are discretized in space by means of a Chebyshev pseudo-spectral collocation method with a two-dimensional mapping implemented in Matlab, which is analogous to that described by \cite{Viola2018a} and \cite{bongarzone2021relaxation}. The numerical scheme requires explicit boundary conditions at $r=0$ in order to regularize the problem on the revolution axis ($r=0$), i.e.
\begin{subequations}
\begin{equation}
\label{eq:BCm0}
m=0:\ \frac{\partial\hat{\eta}}{\partial r}=\frac{\partial\hat{\Phi}}{\partial r}=0,
\end{equation}
\begin{equation}
\label{eq:BCm123}
\textcolor{black}{m}\ge1:\ \hat{\eta}=\hat{\Phi}=0.\ \ \ \ 
\end{equation}
\end{subequations}
We recall the well-known dispersion relation for inviscid gravity waves \citep{Lamb32}, 
\begin{equation}
\label{eq:DispRel}
\omega_{mn}^2=k_{mn}\tanh{\left(k_{mn}H\right)},
\end{equation}
where the wavenumber $k_{mn}$ is given by the \textit{n}th-root of the first derivative of the \textit{m}th-order Bessel function of the first kind satisfying $J'_{m}\left(k_{mn}\right)=0$. By denoting the eigenvector associated with the natural frequency $\omega_{mn}$ as $\hat{\mathbf{q}}_{mn}$, solution of the homogeneous version of equation~\eqref{eq:GenEigProb1} for $\Omega=\omega_{mn}$, it is useful for the rest of the analysis to note that owing to the symmetries of the problem, the system admits the following invariant transformation
\begin{equation}
\label{eq:InvTransf}
\left(\hat{\mathbf{q}}_{mn},+m,\text{i}\omega_{mn}\right)\longrightarrow\left(\hat{\mathbf{q}}_{mn},-m,\text{i}\omega_{mn}\right).
\end{equation}
Such an invariance suggests that the spatial structure, $\hat{\mathbf{q}}\left(r,z\right)$, of the system response to an external forcing with temporal and azimuthal periodicity $\left(\Omega,m\right)$ is the same of that computed for $\left(\Omega,-m\right)$, so that the linear solution form~\eqref{eq:GenEigProb0} holds.

\section{Harmonic single-crest (SC) resonance}\label{sec:Sec4}

With the aim to derive a weakly nonlinear (WNL) system of amplitude equations governing the super-harmonic double-crest dynamics (DC) under longitudinal excitation, we first tackle the simpler problem of harmonic single-crest waves (SC). We look for a third order asymptotic solution of the system
\begin{equation}
\label{eq:WNL_exp}
\mathbf{q}=\left\{\Phi,\eta\right\}^T=\epsilon\mathbf{q}_1+\epsilon^2\mathbf{q}_2+\epsilon^3\mathbf{q}_3+\text{O}\left(\epsilon^4\right),
\end{equation}
where the zero order solution, $\mathbf{q}_0=\mathbf{0}$, associated with the rest state, is omitted.\\
\indent With regards to SC waves and, specifically, to the harmonic response at a driving frequency close to the natural frequency of one of the non-axisymmetric \textcolor{black}{$m=\pm1$} modes, $\omega_{1n}$, we assume here a small forcing amplitude of order $\epsilon^3$. This assumption is justified by the fact that close to resonance, $\Omega\approx\omega_{1n}$, and in absence of dissipation, even a small forcing will induce a large system response. Hence, the analysis is expected to hold for $\Omega=\omega_{1n}+\lambda$, where $\lambda$ is a small detuning parameter assumed of order $\epsilon^2$. In the spirit of the multiple scale technique, we introduce the slow time scale $T_2 = \epsilon^2t$, with $t$ being the fast time scale. Hence, the following scalings are assumed:
\begin{equation}
\label{eq:scaling_SC}
f=\epsilon^3 F,\ \ \ \ \ \Omega=\omega_{1n}+\epsilon^2\Lambda,\ \ \ \ \ T_2=\epsilon^2 t,
\end{equation}
with the auxiliary parameters, $F$ and $\Lambda$, of order $\text{O}\left(1\right)$.\\
\indent Given the azimuthal periodicity of the external forcing, i.e. $m=\pm1$, we assume a leading order solution as the sum of two counter-propagating traveling waves,
\begin{equation}
\label{eq:SC_eps1}
\mathbf{q}_1=A_1\left(T_2\right)\hat{\mathbf{q}}_{1}^{A_1}e^{\text{i}\left(\omega_{1n}t-\theta\right)}+B_1\left(T_2\right)\hat{\mathbf{q}}_{1}^{B_1}e^{\text{i}\left(\omega_{1n}t+\theta\right)}+c.c.,
\end{equation}
where $\hat{\mathbf{q}}_1^{A_1}=\hat{\mathbf{q}}_1^{B_1}$ (owing to~\eqref{eq:InvTransf}) is the eigenmode computed by solving~\eqref{eq:GenEigProb1} for its homogeneous solution at $\Omega=\omega_{1n}$, where $\omega_{1n}$ is given by~\eqref{eq:DispRel}. The complex amplitudes $A_1$ and $B_1$, functions of the slow time scale $T_2$ and still \textcolor{black}{undetermined} at this stage of the expansion, describe the slow time amplitude modulation of the two oscillating waves \textcolor{black}{and must be determined} at a higher order.\\
\indent By pursuing the expansion to the second order, one obtains a linear system forced by combinations of the first order solutions. These forcing terms are proportional to $A_1^2$ and $B_1^2$ (second harmonics), to $|A_1|^2$ and $|B_1|^2$ (steady and axisymmetric mean flow corrections) and to $A_1B_1$ and $A_1\overline{B}_1$ (\textcolor{black}{cross-quadratic} interactions),
\begin{eqnarray}
\label{eq:SC_eps2_forc}
\left(\partial_t\mathsfbi{B}-\textcolor{black}{\mathsfbi{A}_{m}}\right)\mathbf{q}_2=\mathcal{F}_2=\left(|A_1|^2\boldsymbol{\hat{\mathcal{F}}}_2^{A_1\bar{A}_1}+|B_1|^2_1\boldsymbol{\hat{\mathcal{F}}}_2^{B_1\bar{B}_1}\right)\notag\\
+\left(A_1^2\boldsymbol{\hat{\mathcal{F}}}_{2}^{A_1A_1}e^{\text{i}2\left(\omega_{1n}t-\theta\right)}+B_1^2\boldsymbol{\hat{\mathcal{F}}}_{2}^{B_1B_1}e^{\text{i}2\left(\omega_{1n}t+\theta\right)}+c.c.\right)\notag\\
+\left(A_1B_1\boldsymbol{\hat{\mathcal{F}}}_{2}^{A_1B_1}e^{\text{i}2\omega_{1n}t}+A_1\overline{B}_1\boldsymbol{\hat{\mathcal{F}}}_{2}^{A_1\overline{B}_1}e^{-\text{i}2\theta}+c.c.\right).
\end{eqnarray}
\noindent Thus, we seek for a second order solution of the form
\begin{eqnarray}
\label{eq:SC_eps2}
\mathbf{q}_2=|A_1|^2\hat{\mathbf{q}}_2^{A_1\bar{A}_1}+|B_1|^2\hat{\mathbf{q}}_2^{B_1\bar{B}_1}+\left(A_1^2\hat{\mathbf{q}}_{2}^{A_1A_1}e^{\text{i}2\left(\omega_{1n}t-\theta\right)}+B_1^2\hat{\mathbf{q}}_{2}^{B_1B_1}e^{\text{i}2\left(\omega_{1n}t+\theta\right)}+c.c.\right)\notag\\
+\left(A_1B_1\hat{\mathbf{q}}_{2}^{A_1B_1}e^{\text{i}2\omega_{1n}t}+A_1\overline{B}_1\hat{\mathbf{q}}_{2}^{A_1\overline{B}_1}e^{-\text{i}2\theta}+c.c.\right). \ \ \ \ \ 
\end{eqnarray}
Given the invariant transformation~\eqref{eq:InvTransf}, only some of these second order responses need to be computed explicitly, as, e.g., $\hat{\mathbf{q}}_2^{A_1\bar{A}_1}=\hat{\mathbf{q}}_2^{B_1\bar{B}_1}$ and $\hat{\mathbf{q}}_2^{A_1A_1}=\hat{\mathbf{q}}_2^{B_1B_1}$.\\
\indent We now move forward to the $\epsilon^3$--order problem, which is once again a linear problem forced by combinations of \textcolor{black}{the first~\eqref{eq:SC_eps1} and second order solutions~\eqref{eq:SC_eps2}, produced by third order non-linearities such as $\left(\nabla\Phi_1\cdot\nabla\Phi_2+\nabla\Phi_2\cdot\nabla\Phi_1\right)/2$ in the dynamic condition or $\nabla\Phi_1\cdot\nabla\eta_2+\nabla\Phi_2\cdot\nabla\eta_1$ in the kinematic equation,} as well as by the slow time-$T_2$ derivative of the leading order solution and by the external forcing, which was assumed of order $\epsilon^3$,
\begin{eqnarray}
\label{eq:SC_eps3}
\left(\partial_t\mathsfbi{B}-\textcolor{black}{\mathsfbi{A}_{m}}\right)\mathbf{q}_3=\boldsymbol{\mathcal{F}}_3=-\frac{\partial A_1}{\partial T_2} \mathsfbi{B}\hat{\mathbf{q}}_1^{A_1}e^{\text{i}\left(\omega_{1n}t-\theta\right)}-\frac{\partial B_1}{\partial T_2}\mathsfbi{B}\hat{\mathbf{q}}_1^{B_1}e^{\text{i}\left(\omega_{1n}t+\theta\right)}\\
+ |A_1|^2A_1 \boldsymbol{\hat{\mathcal{F}}}_3^{|A_1|^2A_1}e^{\text{i}\left(\omega_{1n}t-\theta\right)}+ |B_1|^2B_1 \boldsymbol{\hat{\mathcal{F}}}_3^{|B_1|^2B_1}e^{\text{i}\left(\omega_{1n}t+\theta\right)}\notag \\
+ |B_1|^2A_1 \boldsymbol{\hat{\mathcal{F}}}_3^{|B_1|^2A_1}e^{\text{i}\left(\omega_{1n}t-\theta\right)}+ |A_1|^2B_1 \boldsymbol{\hat{\mathcal{F}}}_3^{|A_1|^2B_1}e^{\text{i}\left(\omega_{1n}t+\theta\right)}\notag \\
+\frac{1}{2}F\boldsymbol{\hat{\mathcal{F}}}_3^F e^{\text{i}\left(\omega_{1n}t-\theta\right)}e^{\text{i}\Lambda T_2}+\frac{1}{2}F\boldsymbol{\hat{\mathcal{F}}}_3^F e^{\text{i}\left(\omega_{1n}t+\theta\right)}e^{\text{i}\Lambda T_2}\notag \\
+\text{N.R.T.}+c.c.,\nonumber
\end{eqnarray}
with $\boldsymbol{\hat{\mathcal{F}}}_3^F=\left\{0,r/2\right\}^T$ and where $\text{N.R.T.}$ stands for non-resonating terms. These terms are not strictly relevant for further analysis and can therefore be neglected. \textcolor{black}{Amplitudes equations for $A_1$ and $B_1$ are obtained by requiring that secular terms do not appear in the solution to equation~\eqref{eq:SC_eps3}, where secularity} results from all resonant forcing terms in $\boldsymbol{\mathcal{F}}_3$ (see Appendix~D of \cite{bongarzone2022amplitude} for its explicit expression), i.e. all terms sharing the same frequency and wavenumber of $\mathbf{q}_1$, e.g. $\left(\omega,m\right)=\left(\omega_{1n},\pm1\right)$, and in effect all terms explicitly written in~\eqref{eq:SC_eps3}. It follows that a compatibility condition must be enforced through the Fredholm alternative \citep{friedrichs2012spectral}, which imposes the amplitudes $A=\epsilon A_1e^{-\text{i}\lambda t}$ and $B=\epsilon B_1e^{-\text{i}\lambda t}$ to obey the following normal form 
\begin{subequations}
\begin{equation}
\label{eq:AmpEqSCfinalA}
\frac{dA}{dt}=-\text{i}\lambda A + \text{i}\,\frac{\mu_{_{SC}}}{2}f + \text{i}\,\nu_{_{SC}} |A|^2A +\text{i}\,\xi_{_{SC}}|B|^2A,
\end{equation}
\begin{equation}
\label{eq:AmpEqSCfinalB}
\frac{dB}{dt}=-\text{i}\lambda B + \text{i}\,\frac{\mu_{_{SC}}}{2}f + \text{i}\,\nu_{_{SC}} |B|^2B + \text{i}\,\xi_{_{SC}}|A|^2B,
\end{equation}
\end{subequations}
where the physical time $t=T_2/\epsilon^2$ has been reintroduced and where forcing amplitude and detuning parameter are recast in terms of their corresponding physical values, $f=\epsilon^3 F$ and $\lambda=\epsilon^2\Lambda=\Omega-\omega_{1n}$, so as to eliminate the small implicit parameter $\epsilon$ \citep{bongarzone2021impinging,bongarzone2022sub}. The subscript $_{SC}$ stands for single--crest (SC). The various normal form coefficients, which turn out to be real-valued quantities due to the absence of dissipation, are computed as scalar products between the adjoint mode, $\hat{\mathbf{q}}_1^{A_1 \dagger}=\hat{\mathbf{q}}_1^{B_1 \dagger}$, associated with $\hat{\mathbf{q}}_1^{A_1}=\hat{\mathbf{q}}_1^{B_1}$, and the third order resonant forcing terms (see Appendix~\ref{sec:AppB} and \cite{bongarzone2022amplitude} for further details).\\
\indent Once stable stationary solutions are computed, $A$ and $B$ are replaced in expressions~\eqref{eq:SC_eps1} and~\eqref{eq:SC_eps2} and the total harmonic SC wave solution is reconstructed  as
\begin{equation}
\label{eq:SC_sol_reconst}
\mathbf{q}_{SC}=\left\{\Phi,\eta\right\}^T=\epsilon\mathbf{q}_1+\epsilon^2\mathbf{q}_2.
\end{equation}
\indent To this end, it is first convenient to express equations~\eqref{eq:AmpEqSCfinalA}-\eqref{eq:AmpEqSCfinalB} in polar coordinates, i.e. by defining $A=|A|e^{\text{i}\Phi_A}$ and $B=|B|e^{\text{i}\Phi_B}$, and then to introduce the following change of variables, $|a|=|A|+|B|$ and $|b|=|A|-|B|$. By looking for periodic solutions with stationary amplitudes $|A|,\,|B|\ne0$, one can sum and subtract equations~\eqref{eq:AmpEqSCfinalA} and~\eqref{eq:AmpEqSCfinalB}, hence obtaining,
\begin{subequations}
\begin{equation}
\label{eq:AmpEqSCstat_a}
f=a_x\Omega^2=\pm |a|\left(\lambda-\left(\frac{\nu_{_{SC}}+\xi_{SC}}{4}\right)|a|^2-\left(\frac{3\nu_{_{SC}}-\xi_{SC}}{4}\right)|b|^2\right)\frac{1}{\mu_{_{SC}}},
\end{equation}
\begin{equation}
\label{eq:AmpEqSCstat_b}
0=|b|\left(\lambda-\left(\frac{\nu_{_{SC}}+\xi_{_{SC}}}{4}\right)|b|^2-\left(\frac{3\nu_{_{SC}}-\xi_{SC}}{4}\right)|a|^2\right).
\end{equation}
\end{subequations}
\noindent As expected, equation~\eqref{eq:AmpEqSCstat_b} suggests that two possible solutions exist. The planar (or standing) wave solution is retrieved for 
\begin{subequations}
\begin{equation}
\label{eq:planar_harm_b}
|b|=|A|-|B|=0\rightarrow |A|=|B|,
\end{equation}
\begin{equation}
\label{eq:planar_harm_a}
a_x\Omega^2=\pm |a|\left(\lambda-\left(\frac{\nu_{_{SC}}+\xi_{_{SC}}}{4}\right)|a|^2\right)\frac{1}{\mu_{_{SC}}},
\end{equation}
\end{subequations}
whereas the swirling wave solution is found when $|b|\ne0$ and
\begin{subequations}
\begin{equation}
\label{eq:swirling_harm_b}
|b|^2=\left(\lambda-\left(\frac{3\nu_{_{SC}}-\xi_{_{SC}}}{4}\right)|a|^2\right)\left(\frac{4}{\nu_{_{SC}}+\xi_{_{SC}}}\right),
\end{equation}
\begin{equation}
\label{eq:swirling_harm_a}
a_x\Omega^2=\pm 2|a|\left(\frac{\xi_{_{SC}}-\nu_{_{SC}}}{\nu_{_{SC}}+\xi_{_{SC}}}\right)\left(\lambda-\nu_{_{SC}}|a|^2\right)\frac{1}{\mu_{_{SC}}}.
\end{equation}
\end{subequations}
\noindent The various branches prescribed by~\eqref{eq:planar_harm_a} and \eqref{eq:swirling_harm_b}-\eqref{eq:swirling_harm_a} for $|a|$ and $|b|$ as a function of $\tau=\Omega/\omega_{1n}$ and at a fixed non-dimensional shaking amplitude $a_x$ are here computed by means of the Matlab function \textit{fimplicit}.\\
\indent We note that four possible combinations of stationary phases, $\Phi_A$ and $\Phi_B\in\left[0,2\pi\right]$, are in principle admitted, i.e. (i) $\Phi_A=\Phi_B=0$, (ii) $\Phi_A=\Phi_B=\pi$, (iii) $\Phi_A=0$, $\Phi_B=\pi$ and (iv) $\Phi_A=\pi$, $\Phi_B=0$. However, (iii) and (iv) are totally equivalent to (i) and (ii), respectively, with amplitudes $|a|\rightarrow|b|$ and $|b|\rightarrow|a|$. Therefore, only combinations (i) $\Phi_A=\Phi_B=\Phi=0$ and (ii) $\Phi=\pi$, which produce the $\pm$ sign in~\eqref{eq:AmpEqSCstat_a}, are retained.

%\subsection{Comparison with existing experiments and theoretical predictions}\label{subsec:Sec4sec1}

\begin{figure}
%\centering
\hspace*{1.9cm}\includegraphics[width=0.6\textwidth]{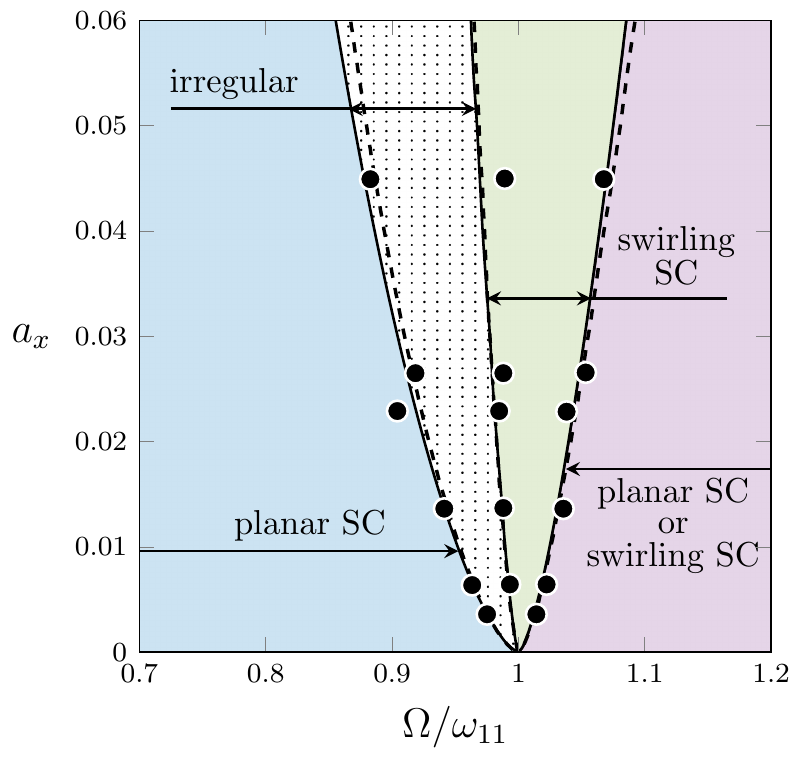}
\caption{Estimates of bounds, in the $\left(\Omega/\omega_{11},a_x\right)$-plane, between the frequency ranges where planar, irregular and swirling waves occur when the container undergoes a longitudinal and harmonic motion. Filled markers: experiments by \cite{royon2007liquid}. Black dashed lines: theoretical prediction by \cite{faltinsen2016resonant}, whose theoretical curves have been here reproduced by manually sampling those reported in their original figure~8($a$).}
\label{fig:Fig4} 
\end{figure}

\begin{figure}
\centering
\includegraphics[width=0.925\textwidth]{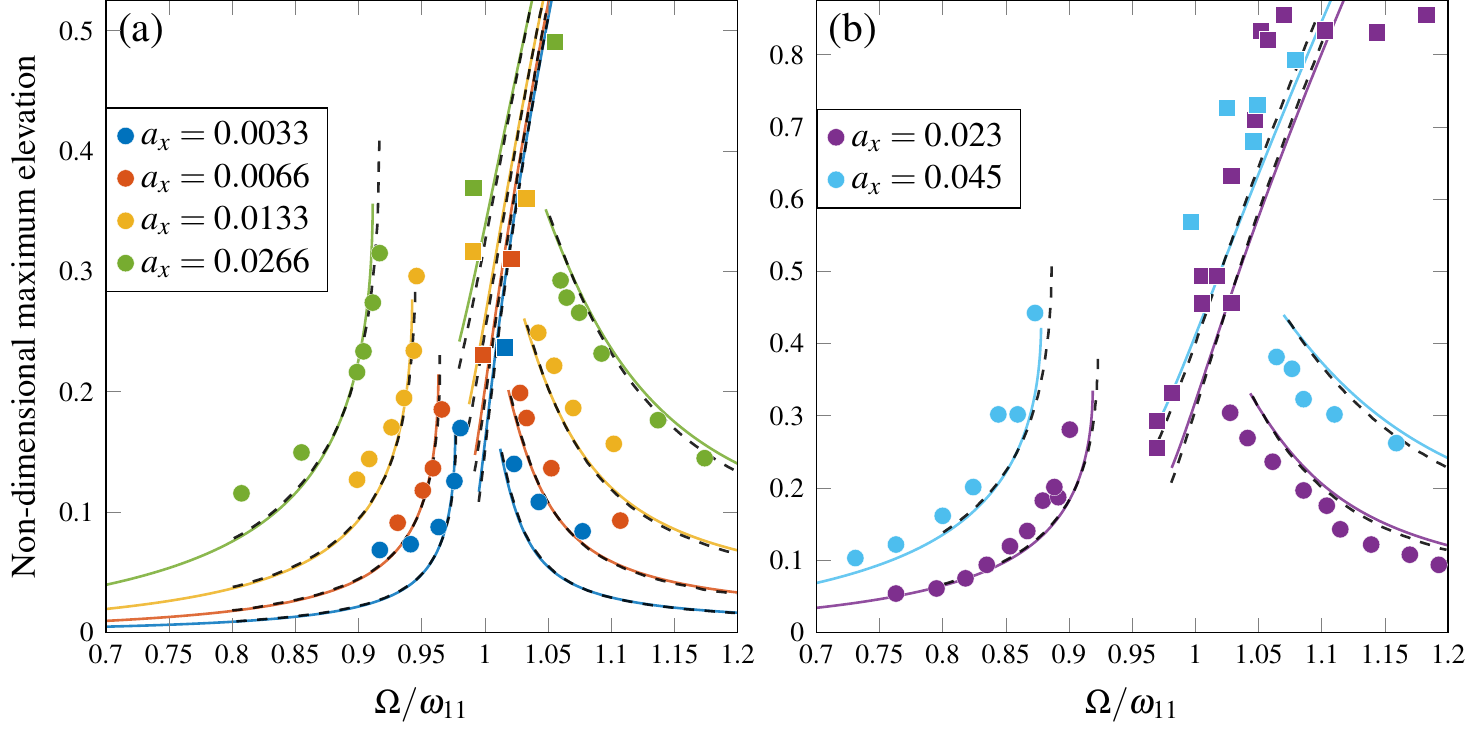}
\caption{Non-dimensional maximum steady-state wave elevation, $\max_{t,\theta=0,\pi/2}\,\eta$,  (the maximum is taken from values at two probes located at $\left(x,y\right)=\left(0.875,0\right)$ and $\left(0,0.875\right)$) versus the forcing frequency $\Omega/\omega_{11}$ and for different $x$-longitudinal shaking amplitudes, $a_x$: ($a$) $0.0033$, $0.0066$, $0.0133$ and $0.0266$; ($b$) $0.023$ and $0.045$. Markers are associated with two experimental series by \cite{royon2007liquid} (experimental data from their original figures 2 (now ($a$)) and 7 (now ($b$)). Filled circles correspond to measurements done for the planar regime, whereas filled squares indicate swirling. The black dashed lines represent the stable branches predicted by \cite{faltinsen2016resonant}. Their curves have been here carefully reproduced by manually sampling those reported in their original figure~10 in the range of frequency available, i.e. $\Omega/\omega_{11}\in\left[0.7,1.2\right]$. Colored solid lines correspond to the present theoretical predictions for stable branches.}
\label{fig:Fig3} 
\end{figure}

In figure~\ref{fig:Fig4} we reproduce figure~8 of \cite{faltinsen2016resonant}, which shows the estimates of bounds between the frequency ranges where harmonic planar, irregular and swirling waves occur. The outcomes of the present analysis are consistent with those of \cite{faltinsen2016resonant} and with the experimental measurements by \cite{royon2007liquid}. The values of the normal form coefficients $\mu_{_{SC}}$, $\nu_{_{SC}}$ and $\xi_{_{SC}}$ reported in table~\ref{tab:Tab3} \textcolor{black}{of Appendix~\ref{sec:AppB}} confirm that the stability boundaries vary  weakly with the liquid depth, as stated by \cite{faltinsen2016resonant} for non-dimensional fluid depths \textcolor{black}{$H\gtrsim1.05$}, but strongly depend on the forcing amplitude, with the frequency range for irregular and swirling waves widening for increasing forcing amplitudes. In this context, irregular means that both the planar and the swirling wave solutions are unstable, hence one could expect irregular and chaotic patterns with a switching between planar and swirling motion. The green shaded region corresponds to stable single-crest (SC) swirling waves, while the light purple shaded region corresponds to the multi-solution regime, where both stable swirling SC and planar SC wave motions are possible depending on the initial transient, i.e. on the initial conditions, as typical of hysteretic systems.\\
\indent In figure~\ref{fig:Fig3}(a) and (b) the non-dimensional maximum steady-state wave elevation, computed by reconstructing the total flow solution in accordance with~\eqref{eq:SC_sol_reconst}, is compared with the theoretical estimations by \cite{faltinsen2016resonant} (black dashed lines) from their figure~8 and with the corresponding experimental measurements by \cite{royon2007liquid} (colored filled markers). The agreement between the present model and experiments is fairly good and consistent with predictions by \cite{faltinsen2016resonant}. The larger disagreement between theory and experiments at smaller forcing amplitudes was tentatively attributed by \cite{faltinsen2016resonant} to the fact that the actual elevation of these wave amplitudes was approximately $1\,\text{mm}$ and may therefore be more difficult to measure with sufficient accuracy. A comparable mismatch is here retrieved.\\
\indent As a side comment, we note that, within the present inviscid framework, the lower left stable planar branch, $\Omega/\omega_{11}<1$, is obtained for a phase $\Phi=0$, which implies a fluid motion 

%\clearpage


%\clearpage
\noindent in phase with the container motion, whereas the lower right planar branch, $\Omega/\omega_{11}>1$, has a phase $\Phi=\pi$, hence implying a phase opposition. The stable swirling branch is characterized by $\Phi=0$. This is consistent with previous studies \citep{royon2007liquid}.

\section{Super-harmonic double-crest (DC) resonance}\label{sec:Sec5}

We now tackle the double--crest (DC) wave response to longitudinal shaking, whose investigation represents the core of the present work. We remind that the double-crest dynamics occurs at a driving frequency $\Omega\approx\omega_{2n}/2$ (see figure~4 of \cite{reclari2014surface}). For the sake of generality, the following analysis is therefore formalized for any mode $\left(2,n\right)$, i.e. $\Omega=\omega_{2n}/2+\lambda$, where $\lambda$ is the small detuning parameter.\\
\indent By analogy with \cite{bongarzone2022amplitude}, the leading order solution is here assumed to be given by the sum of a particular solution, given by the linear response to the external forcing, computed by solving~\eqref{eq:GenEigProb1} with $\Omega=\omega_{2n}/2$ and $m=\pm1$, and a homogeneous solution, represented by the two natural modes for \textcolor{black}{$\left(m,n\right)=\left(\pm2,n\right)$} associated with $\omega_{2n}$, up to their amplitudes to be determined at higher orders. At second order, quadratic terms in $\left(\Omega,m\right)=\left(\omega_{2n}/2,\pm 1\right)$ will produce resonant terms in $\left(\omega_{2n},\pm2\right)$. These $\epsilon^2$--order resonating terms will then require, in the spirit of multiple timescale analysis, an additional second order solvability condition, hence suggesting that two slow time scales exist, namely $T_1$ and $T_2$. Thus, the asymptotic scalings of the weakly nonlinear expansion for double-crest (DC) waves are the following:
\begin{equation}
\label{eq:scaling_DC}
f=\epsilon F,\ \ \ \ \ \Omega=\omega_{2n}/2+\epsilon\Lambda,\ \ \ \ \ T_1=\epsilon t,\ \ \ \ \ T_2=\epsilon^2 t,
\end{equation}
\noindent with a first order solution reading
\begin{eqnarray}
\label{eq:DC_sol_eps1}
\mathbf{q}_{1}=A_2\left(T_1,T_2\right)\hat{\mathbf{q}}_1^{A_2}e^{\text{i}\left(\omega_{2n}t-2\theta\right)}+B_2\left(T_1,T_2\right)\hat{\mathbf{q}}_1^{B_2}e^{\text{i}\left(\omega_{2n}t+2\theta\right)}\notag \\
+\frac{1}{2}F\hat{\mathbf{q}}_1^{F}e^{\text{i}\left(\left(\omega_{2n}/2\right)t-\theta\right)}e^{\text{i}\Lambda T_1}+\frac{1}{2}F\hat{\mathbf{q}}_1^{F}e^{\text{i}\left(\left(\omega_{2n}/2\right)t+\theta\right)}e^{\text{i}\Lambda T_1}+c.c.\,.
\end{eqnarray}
\begin{figure}
\centering
%\hspace*{1.9cm}
\includegraphics[width=0.8\textwidth]{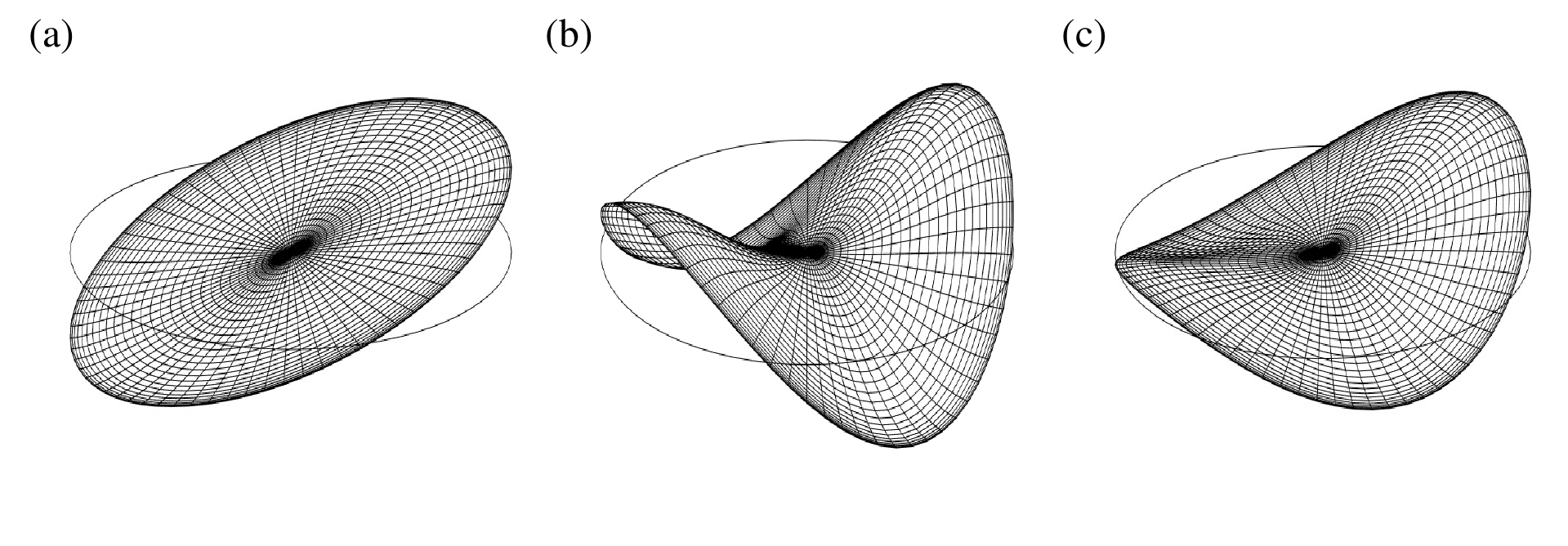}
\caption{Spatial structures of the first order contributions (a) $\mathbf{q}_1^F\left(r,z\right)e^{\text{i}}\cos{\theta}$ (single-crest SC) and (b) $\hat{\mathbf{q}}_1^{A_2}\cos{2\theta}=\hat{\mathbf{q}}_1^{B_2}\cos{2\theta}$ (double-crest DC) appearing in~\eqref{eq:DC_sol_eps1} and computed for $t=0$ and $T_1=0$. (c) Superposition of (a) and (b). Here the corresponding amplitudes have been arbitrarily chosen for visualization purposes,but we note that, while amplitude $A_2$ and $B_2$ still need to be determined, the amplitude of the single-crest solution (a) is univocally defined once the amplitude, $F$, and the oscillation frequency, $\Omega$, of the external driving are prescribed.}
\label{fig:Fig_ModeDC} 
\end{figure}
\noindent In~\eqref{eq:DC_sol_eps1}, $\hat{\mathbf{q}}_1^{A_2}=\hat{\mathbf{q}}_1^{B_2}$, whereas $A_2$ and $B_2$ are the unknown slow time amplitude modulations, here functions of the two time scales $T_1$ and $T_2$. The second order linearized forced problem reads
\begin{eqnarray}
\label{eq:DC_eps2}
\left(\partial_t\mathsfbi{B}-\textcolor{black}{\mathsfbi{A}_{m}}\right)\mathbf{q}_2=\boldsymbol{\mathcal{F}}_2=\boldsymbol{\mathcal{F}}_2^{ij}-\left(\frac{\partial A_2}{\partial T_1}\mathsfbi{B}\hat{\mathbf{q}}_1^{A_2}e^{\text{i}\left(\omega_{2n}t-2\theta\right)} + \frac{\partial B_2}{\partial T_1}\mathsfbi{B}\hat{\mathbf{q}}_1^{B_2}e^{\text{i}\left(\omega_{2n}t+2\theta\right)}+c.c.\right)\notag \\ 
-\text{i}\Lambda F\left( \frac{1}{2} \mathsfbi{B}\hat{\mathbf{q}}_1^{F}e^{\text{i}\left(\left(\omega_{2n}/2\right)t-\theta\right)}\textcolor{black}{e^{\text{i}\Lambda T_1}} +\frac{1}{2} \mathsfbi{B}\hat{\mathbf{q}}_1^{F}e^{\text{i}\left(\left(\omega_{2n}/2\right)t+\theta\right)}\textcolor{black}{e^{\text{i}\Lambda T_1}}+c.c.\right).\ \ \ \ \  
\end{eqnarray}
\noindent The first order solution is indeed made of \textcolor{black}{8} different contributions (including the complex conjugates) and it generates, in total, 36 different second order forcing terms, \textcolor{black}{here implicitly gathered in $\boldsymbol{\mathcal{F}}_2^{ij}$, each characterized by a certain oscillation frequency and azimuthal periodicity. For the sake of brevity, indices $\left(i,j\right)$ are used to remind that each forcing is proportional to a quadratic combination of leading order amplitudes. For instance, the quadratic interaction of $A_2\left(T_1,T_2\right)\hat{\mathbf{q}}_1^{A_2}e^{\text{i}\left(\omega_{2n}t-2\theta\right)}$ with itself will have indices $\left(i=A_2,j=A_2\right)$ and will produce a forcing term proportional to $A_2^2$, i.e $\boldsymbol{\mathcal{F}}_2^{A_2A_2}$.} The additional eight forcing terms, with their complex conjugates, appearing in~\eqref{eq:DC_eps2} stem from the time derivative of the first-order solution~\eqref{eq:DC_sol_eps1} with respect to the first-order slow time scale $T_1$. None of the forcing terms in~\eqref{eq:DC_eps2} is resonant, as their oscillation frequency \textcolor{black}{or} azimuthal wavenumber differ from those of the leading order homogeneous solution, except the two terms produced by the second--harmonic of the leading order particular solution, i.e. $\boldsymbol{\mathcal{F}}_2^{FF}=\textcolor{black}{\frac{1}{4}}F^2\boldsymbol{\hat{\mathcal{F}}}_2^{FF}e^{\text{i}\left(\omega_{2n}t-2\theta\right)}e^{\text{i}2\Lambda T_1}+\textcolor{black}{\frac{1}{4}}F^2\boldsymbol{\hat{\mathcal{F}}}_2^{FF}e^{\text{i}\left(\omega_{2n}t+2\theta\right)}e^{\text{i}2\Lambda T_1}+c.c.\,$. To avoid secular terms, a second order compatibility condition is thus imposed, requiring that the following normal form equations are verified
\begin{equation}
\label{eq:DC_AmpEq_eps2_AB}
\frac{\partial A_2}{\partial T_1}=\text{i}\,\frac{\mu_{_{DC}}}{4} F^2e^{\text{i}2\Lambda T_1},\ \ \ \ \ \frac{\partial B_2}{\partial T_1}=\text{i}\,\frac{\mu_{_{DC}}}{4} F^2e^{\text{i}2\Lambda T_1}.
\end{equation}
\noindent Taken alone, the dynamics resulting from system~\eqref{eq:DC_AmpEq_eps2_AB} is still of little relevance, \textcolor{black}{since it can be shown that the wave amplitudes $A_2$ and $B_2$ scale like $\sim \frac{1}{\Lambda}$, hence diverging symmetrically to infinity for $\Lambda\rightarrow 0$ ($\Omega\rightarrow\omega_{2n}/2$)} in absence of any restoring term, i.e. the nonlinear mechanism responsible for the finite amplitude saturation, which only comes into play at order $\epsilon^3$. The expansion must be therefore pursued up to the next order, and thereby one must solve for the second-order solution \citep{fujimura1989equivalence,fujimura1991methods}.\\
\indent By substituting~\eqref{eq:DC_sol_eps1} and~\eqref{eq:DC_AmpEq_eps2_AB} in the forcing expression, equation~\eqref{eq:DC_eps2} can be rewritten as
\begin{eqnarray}
\label{eq:DC_eps2_1}
\left(\partial_t\mathsfbi{B}-\textcolor{black}{\mathsfbi{A}_{m}}\right)\mathbf{q}_2= \boldsymbol{\mathcal{F}}_{2_{NRT}}^{ij}+\boldsymbol{\mathcal{F}}_{2_{RT}}^{ij}=\boldsymbol{\mathcal{F}}_{2_{NRT}}^{ij}+c.c.+\\
\frac{1}{4}F^2\left(\boldsymbol{\hat{\mathcal{F}}}_2^{FF}-\text{i}\,\mu_{_{DC}} \mathsfbi{B}\hat{\mathbf{q}}_1^{A_2}\right)e^{\text{i}\left(\omega_{2n}t-2\theta\right)}e^{\text{i}2\Lambda T_1}+c.c.+\notag \\
\frac{1}{4}F^2\left(\boldsymbol{\hat{\mathcal{F}}}_2^{FF}-\text{i}\,\mu_{_{DC}} \mathsfbi{B}\hat{\mathbf{q}}_1^{B_2}\right)e^{\text{i}\left(\omega_{2n}t+2\theta\right)}e^{\text{i}2\Lambda T_1}+c.c.,\ \ \ \notag
\end{eqnarray}
\noindent where the subscripts $_{NRT}$ and $_{RT}$ denote non-resonating and resonating terms, respectively. Note that the term proportional to $\Lambda F$ in~\eqref{eq:DC_eps2} has been included in the non-resonating forcing terms, while resonant terms are written explicitly. The compatibility condition is now satisfied, meaning that the new resonant forcing term is orthogonal to the adjoint mode, $\hat{\mathbf{q}}_1^{A_2\dagger}=\overline{\hat{\mathbf{q}}}_1^{A_2}$, by construction so that, according to the Fredholm alternative, a non-trivial \textcolor{black}{unique} solution can be computed. Hence, \textcolor{black}{we can write the second order solution as}
\vspace{-0.5cm}
\begin{eqnarray}
\label{eq:DC_eps2_full_sol}
\mathbf{q}_2=\left(|A_2|^2\hat{\mathbf{q}}_2^{A_2\bar{A}_2}+\frac{1}{4}|F|^2\hat{\mathbf{q}}_2^{F\bar{F}}\right)+ \ \\
\left(A_2^2\hat{\mathbf{q}}_2^{A_2A_2}e^{\text{i}\left(2\omega_{2n}t-4\theta\right)}+\frac{1}{2}\Lambda F\hat{\mathbf{q}}_2^{\Lambda F}e^{\text{i}\left(\left(\omega_{2n}/2\right)t-\theta\right)}e^{\text{i}\Lambda T_1}+c.c.\right)+\ \notag\\
\left(\frac{1}{2}A_2F\hat{\mathbf{q}}_2^{A_2F}e^{\text{i}\left(\left(3\omega_{2n}/2\right)t-3\theta\right)}e^{\text{i}\Lambda T_1}+\frac{1}{2}A_2\overline{F}\hat{\mathbf{q}}_2^{A_2\overline{F}}e^{\text{i}\left(\left(\omega_{2n}/2\right)t-\theta\right)}e^{-\text{i}\Lambda T_1}+c.c.\right)+\ \notag\\
\left(|B_2|^2\hat{\mathbf{q}}_2^{B_2\bar{B}_2}+\frac{1}{4} |F|^2\hat{\mathbf{q}}_2^{F\bar{F}}\right)+\ \notag\\
\left(B_2^2\hat{\mathbf{q}}_2^{B_2B_2}e^{\text{i}\left(2\omega_{2n}t+4\theta\right)}+\frac{1}{2}\Lambda F\hat{\mathbf{q}}_2^{\Lambda F}e^{\text{i}\left(\left(\omega_{2n}/2\right)t+\theta\right)}e^{\text{i}\Lambda T_1}+c.c.\right)+\ \notag\\
\left(\frac{1}{2} B_2F\hat{\mathbf{q}}_2^{B_2F}e^{\text{i}\left(\left(3\omega_{2n}/2\right)t+3\theta\right)}e^{\text{i}\Lambda T_1}+\frac{1}{2} B_2\overline{F}\hat{\mathbf{q}}_2^{B_2\overline{F}}e^{\text{i}\left(\left(\omega_{2n}/2\right)t+\theta\right)}e^{-\text{i}\Lambda T_1}+c.c.\right)+\ \notag \\
\left(A_2B_2\hat{\mathbf{q}}_2^{A_2B_2}e^{\text{i}2\omega_{2n}t}+A_2\overline{B}_2\hat{\mathbf{q}}_2^{A_2\overline{B}_2}e^{-\text{i}4\theta}+c.c.\right)+\ \notag\\
\left(\frac{1}{4}F^2\hat{\mathbf{q}}_2^{FF}e^{\text{i}\omega_{2n}t}e^{\text{i}2\Lambda T_1}+\frac{1}{4}F\overline{F}\hat{\mathbf{q}}_2^{F\overline{F}}e^{-\text{i}2\theta}+c.c.\right)+\ \notag\\
\left(\frac{1}{2}A_2F\hat{\mathbf{q}}_2^{A_2F}e^{\text{i}\left(\left(3\omega_{2n}/2\right)t-\theta\right)}e^{\text{i}\Lambda T_1}+\frac{1}{2}A_2\overline{F}\hat{\mathbf{q}}_2^{A_2\overline{F}}e^{\text{i}\left(\left(\omega_{2n}/2\right)t-3\theta\right)}e^{-\text{i}\Lambda T_1}+c.c.\right)+\ \notag\\
\left(\frac{1}{2}B_2F\hat{\mathbf{q}}_2^{B_2F}e^{\text{i}\left(\left(3\omega_{2n}/2\right)t+\theta\right)}e^{\text{i}\Lambda T_1}+\frac{1}{2}B_2\overline{F}\hat{\mathbf{q}}_2^{B_2\overline{F}}e^{\text{i}\left(\left(\omega_{2n}/2\right)t+3\theta\right)}e^{-\text{i}\Lambda T_1}+c.c.\right)+\ \notag \\
\left(\frac{1}{4} F^2\hat{\mathbf{q}}_2^{FF}e^{\text{i}\left(\omega_{2n}t-2\theta\right)}e^{\text{i}2\Lambda T_1}+\frac{1}{4} F^2\hat{\mathbf{q}}_2^{FF}e^{\text{i}\left(\omega_{2n}t+2\theta\right)}e^{\text{i}2\Lambda T_1}+c.c.\right).\ \ \ \notag
\end{eqnarray}
\noindent All non-resonant responses in~\eqref{eq:DC_eps2_full_sol} are handled similarly, i.e. they are computed in Matlab by performing a simple matrix inversion using standard LU solvers. As anticipated above, although the operator associated with the resonant forcing term, i.e. $\left(\text{i}\omega_{2n}\mathsfbi{B}-\mathsfbi{A}_{2}\right)$, is singular, the value of the normal form coefficient $\mu_{_{DC}}$ ensures that a non-trivial solution for $\hat{\mathbf{q}}_2^{F^2}$ exists. Diverse approaches can be followed to compute this response, which was here computed by using the \textit{pseudo-inverse} matrix of the singular operator \citep{orchini2016weakly}. We also recall that due to the invariant transformation~\eqref{eq:InvTransf} only some of the spatial structures appearing in~\eqref{eq:DC_eps2_full_sol} need to be computed. Lastly, at third order in $\epsilon$, the problem reads
\begin{eqnarray}
\label{eq:DC_eps3}
\left(\partial_t\mathsfbi{B}-\textcolor{black}{\mathsfbi{A}_{m}}\right)\mathbf{q}_3=\boldsymbol{\mathcal{F}}_3\\
=-\frac{\partial A_2}{\partial T_2 }\mathsfbi{B}\hat{\mathbf{q}}_1^{A_2}e^{\text{i}\left(\omega_{2n}t-2\theta\right)}-\frac{\partial B_2}{\partial T_2 }\mathsfbi{B}\hat{\mathbf{q}}_1^{B_2}e^{\text{i}\left(\omega_{2n}t+2\theta\right)}\notag\\ 
- \text{i}\frac{1}{4}2\Lambda F^2\mathsfbi{B}\hat{\mathbf{q}}_2^{F^2}e^{\text{i}\left(\omega_{2n}t-2\theta\right)}e^{\text{i}2\Lambda T_1} - \text{i}\frac{1}{4}2\Lambda F^2\mathsfbi{B}\hat{\mathbf{q}}_2^{F^2}e^{\text{i}\left(\omega_{2n}t+2\theta\right)}e^{\text{i}2\Lambda T_1}\notag\\
+ |A_2|^2A_2 \boldsymbol{\hat{\mathcal{F}}}_3^{|A_2|^2A_2}e^{\text{i}\left(\omega_{2n}t-2\theta\right)}+ |B_2|^2B_2 \boldsymbol{\hat{\mathcal{F}}}_3^{|B_2|^2B_2}e^{\text{i}\left(\omega_{2n}t+2\theta\right)}\notag\\
+ |B_2|^2A_2 \boldsymbol{\hat{\mathcal{F}}}_3^{|B_2|^2A_2}e^{\text{i}\left(\omega_{2n}t-2\theta\right)}+ |A_2|^2B_2 \boldsymbol{\hat{\mathcal{F}}}_3^{|A_2|^2B_2}e^{\text{i}\left(\omega_{2n}t+2\theta\right)}\notag\\
+\frac{1}{4}F^2A_2\boldsymbol{\hat{\mathcal{F}}}_3^{|F|^2A_2}e^{\text{i}\left(\omega_{2n}t-2\theta\right)}+\frac{1}{4}F^2B_2\boldsymbol{\hat{\mathcal{F}}}_3^{|F|^2B_2}e^{\text{i}\left(\omega_{2n}t+2\theta\right)}\notag\\
+\frac{1}{4}\Lambda F^2 \boldsymbol{\hat{\mathcal{F}}}_3^{\Lambda F^2} e^{\text{i}\left(\omega_{2n}t-2\theta\right)}e^{\text{i}2\Lambda T_1}+\frac{1}{4}\Lambda F^2 \boldsymbol{\hat{\mathcal{F}}}_3^{\Lambda F^2} e^{\text{i}\left(\omega_{2n}t+2\theta\right)}e^{\text{i}2\Lambda T_1}+\text{N.R.T.}+c.c.\,,\notag
\end{eqnarray}
where the first two forcing terms arise from the time-derivative of the first order solution with respect to the second order slow time scale $T_2$ and from that of the second order solution with respect to the first order slow time scale $T_1$, respectively (see Appendix~D of \cite{bongarzone2022amplitude} for the full expression of $\boldsymbol{\mathcal{F}}_2$ and $\boldsymbol{\mathcal{F}}_3$). Once again, all terms explicitly written in~\eqref{eq:DC_eps3} are resonant, as they share the same pair $\left(\omega_{2n},\pm2\right)$ than the first order homogeneous solutions, hence a third order compatibility condition, leading to the following normal form, must be enforced
\begin{subequations}
\begin{equation}
\label{eq:AmpEqDCfinal0_A}
\frac{\partial A_2}{\partial T_2}=\text{i}\,\frac{\zeta_{_{DC}}}{4} \Lambda F^2e^{\text{i}2\Lambda T_1}+\text{i}\,\frac{\chi_{_{DC}}}{4}A_2F^2+\text{i}\,\nu_{_{DC}} |A_2|^2A_2+\text{i}\,\xi_{_{DC}}|B_2|^2A_2,
\end{equation}
\begin{equation}
\label{eq:AmpEqDCfinal0_B}
\frac{\partial B_2}{\partial T_2}=\text{i}\,\frac{\zeta_{_{DC}}}{4} \Lambda F^2e^{\text{i}2\Lambda T_1}+\text{i}\,\frac{\chi_{_{DC}}}{4}B_2F^2+\text{i}\,\nu_{_{DC}} |B_2|^2B_2+\text{i}\,\xi_{_{DC}}|A_2|^2B_2.
\end{equation}
\end{subequations}
\noindent \textcolor{black}{where the coefficients are defined in Appendix~\ref{sec:AppB}.}\\
\indent As a last step in the derivation of the final amplitude equation for the double--crest (DC) waves and in order to eliminate the implicit small parameter $\epsilon$, we unify systems~\eqref{eq:DC_AmpEq_eps2_AB} and~\eqref{eq:AmpEqDCfinal0_A}-\eqref{eq:AmpEqDCfinal0_B} into a single system of equations recast in terms of the physical time $t=T_1/\epsilon=T_2/\epsilon^2$, physical forcing control parameters, $f=\epsilon F$, $\lambda=\epsilon\Lambda$ and total amplitudes, $A=\epsilon A_2e^{-\text{i}2\lambda t}$ and $B=\epsilon B_2e^{-\text{i}2\lambda t}$. This is achieved by summing~\eqref{eq:DC_AmpEq_eps2_AB} to~\eqref{eq:AmpEqDCfinal0_A} and~\eqref{eq:AmpEqDCfinal0_B} along with their respective weights $\epsilon^2$ and $\epsilon^3$, thus obtaining
\begin{subequations}
\begin{equation}
\label{eq:AmpEqDCfinal_A}
\frac{dA}{dt}=-\text{i}\,\left(2\lambda-\frac{\chi_{_{DC}}}{4} f^2\right) A + \text{i}\,\frac{\left(\zeta_{_{DC}}\lambda + \mu_{_{DC}}\right)}{4}  f^2 + \text{i}\,\nu_{_{DC}} |A|^2A+\text{i}\,\xi_{_{DC}}|B|^2A,
\end{equation}
\begin{equation}
\label{eq:AmpEqDCfinal_B}
\frac{dB}{dt}=-\text{i}\,\left(2\lambda-\frac{\chi_{_{DC}}}{4} f^2\right) B + \text{i}\,\frac{\left(\zeta_{_{DC}}\lambda + \mu_{_{DC}}\right)}{4} f^2 + \text{i}\,\nu_{_{DC}} |B|^2B+\text{i}\,\xi_{_{DC}}|A|^2B.
\end{equation}
\end{subequations}
\indent \textcolor{black}{We note that no second order homogeneous solutions, e.g. proportional to amplitudes $C_2\left(T_1,T_2\right)$ and $D_2\left(T_1,T_2\right)$, have been accounted for in~\eqref{eq:DC_eps2_full_sol}, as their presence will produce two resonant third order terms, $\frac{\partial C_2}{\partial T_1}\mathsfbi{B}\hat{\mathbf{q}}_2^{C_2}e^{\text{i}\left(\omega_{2n}t-2\theta\right)}$ ($\hat{\mathbf{q}}_2^{C_2}=\hat{\mathbf{q}}_2^{A_2}$) and $\frac{\partial D_2}{\partial T_1}\mathsfbi{B}\hat{\mathbf{q}}_2^{D_2}e^{\text{i}\left(\omega_{2n}t+2\theta\right)}$ ($\hat{\mathbf{q}}_2^{C_2}=\hat{\mathbf{q}}_2^{A_2}$), that can be incorporated in the final amplitude equations~\eqref{eq:AmpEqDCfinal_A}-\eqref{eq:AmpEqDCfinal_B} by simply defining $A=\epsilon\left( A_2+\epsilon C_2\right)e^{-\text{i}2\lambda t}$ and $B=\epsilon\left( B_2+\epsilon D_2\right)e^{-\text{i}2\lambda t}$.}\\
\indent As in \S\ref{sec:Sec4}, we first turn to polar coordinates, $A=|A|e^{\text{i}\Phi_A}$ and $B=|B|e^{\text{i}\Phi_A}$, and we split the modulus and phase parts of~\eqref{eq:AmpEqDCfinal_A}-\eqref{eq:AmpEqDCfinal_B}. We then look for stationary solutions, $d/dt=0$ with $|A|,\,|B|\ne0$ \textcolor{black}{($\Phi_A=\Phi_B=\Phi=0,\pi$, see \S~\ref{sec:Sec4})}. By summing and subtracting \eqref{eq:AmpEqDCfinal_A} and~\eqref{eq:AmpEqDCfinal_B}, after introducing the auxiliary amplitudes $|a|=|A|+|B|$ and $|b|=|A|-|B|$, the following implicit relations are obtained,
\begin{subequations}
\begin{equation}
\label{eq:imp_rel_a}
f^2=|a|\left(2\lambda-\frac{\nu_{_{DC}}+\xi_{_{DC}}}{4}|a|^2-\frac{3\nu_{_{DC}}-\xi_{_{DC}}}{4}|b|^2\right)\frac{4}{\left(|a|\chi_{_{DC}}\pm2\left(\zeta_{_{DC}}\lambda+\mu_{_{DC}}\right)\right)},
\end{equation}
\begin{equation}
\label{eq:imp_rel_b}
0=|b|\left(\frac{\chi_{_{DC}}}{4}f^2-\left(2\lambda-\frac{\nu_{_{DC}}+\xi_{_{DC}}}{4}|b|^2-\frac{3\nu_{_{DC}}-\xi_{_{DC}}}{4}|a|^2\right)\right),
\end{equation}
\end{subequations}
\noindent with $f=a_x\Omega^2$ and $\lambda=\Omega-\omega_{2n}/2$. By analogy with harmonic forcing conditions, two possible (super-harmonic) solutions exist, i.e. a planar wave solution for $|b|=0$,
\begin{equation}
\label{eq:planar_DC_sol}
f=\sqrt{|a|\left(2\lambda-\frac{\nu_{_{DC}}+\xi_{_{DC}}}{4}|a|^2\right)\frac{4}{\left(|a|\chi_{_{DC}}\pm2\left(\zeta_{_{DC}}\lambda+\mu_{_{DC}}\right)\right)}},
\end{equation}
\noindent and a swirling solution for $|b|\ne0$ \textcolor{black}{defined by},
\begin{subequations}
\begin{equation}
\label{eq:swirl_DC_sol_b}
|b|^2=\left(2\lambda-\frac{\chi_{_{DC}}}{4}f^2-\frac{3\nu_{_{DC}}-\xi_{_{DC}}}{4}|a|^2\right)\left(\frac{4}{\nu_{_{DC}}+\xi_{_{DC}}}\right),
\end{equation}
\begin{equation}
\label{eq:swirl_DC_sol_a}
f=\sqrt{2|a|\left(\frac{\xi_{_{DC}}-\nu_{_{DC}}}{\nu_{_{DC}}+\xi_{_{DC}}}\right)\left(2\lambda-\nu_{_{DC}}|a|^2\right)\frac{4}{\left(2|a|\frac{\left(\xi_{_{DC}}-\nu_{_{DC}}\right)}{\left(\nu_{_{DC}}+\xi_{_{DC}}\right)}\chi_{_{DC}}\pm2\left(\zeta_{_{DC}}\lambda+\mu_{_{DC}}\right)\right)}},
\end{equation}
\end{subequations}
\noindent where only real solutions corresponding to $f=a_x\Omega^2>0$ are retained, as the combinations $a_x\Omega^2<0$ are not physically meaningful.

\textcolor{black}{The stability of such stationary solutions $\mathbf{y}_{s}=\left(|A|,\Phi_A,|B|,\Phi_B\right)$ is computed by introducing small amplitude and phase perturbations ($\ll 1$) with the ansatz $\mathbf{y}_{p}\left(t\right)=\left(|A_p|,\Phi_{A,p},|B_p|,\Phi_{B,p}\right)e^{st}$ in~\eqref{eq:AmpEqDCfinal_A}-\eqref{eq:AmpEqDCfinal_B}, which are then linearized around $\mathbf{y}_0$, hence obtaining at first order an eigenvalue problem in the complex eigenvalue $s=s_R+\text{i}s_I$. For each $\left(|A|,\Phi_A,|B|,\Phi_B\right)$ one obtains four eigenvalues $s$ and if the real part $s_R$ of at least one of these eigenvalue is positive, then that configuration is deemed as unstable. An analogous procedure has been followed for the case of harmonic resonances discussed in \S\ref{sec:Sec4}.}

Once the various branches for $|a|$ and $|b|$ as a function of $\tau=\Omega/\omega_{2n}$ and at a fixed non-dimensional shaking amplitude $a_x$ are computed and their stability is determined, amplitudes $A$ and $B$ are substituted in~\eqref{eq:DC_sol_eps1} and~\eqref{eq:DC_eps2_full_sol}, so that the total flow solution predicted by the WNL for DC waves is reconstructed as
\begin{equation}
\label{eq:DC_sol_reconst}
\mathbf{q}_{DC}=\left\{\Phi,\eta\right\}^T=\epsilon\mathbf{q}_1+\epsilon^2\mathbf{q}_2.
\end{equation}

\subsection{Branching diagrams and super-harmonic stability chart}\label{subsec:Sec5sub1}

\begin{figure}
\centering
\includegraphics[width=0.999\textwidth]{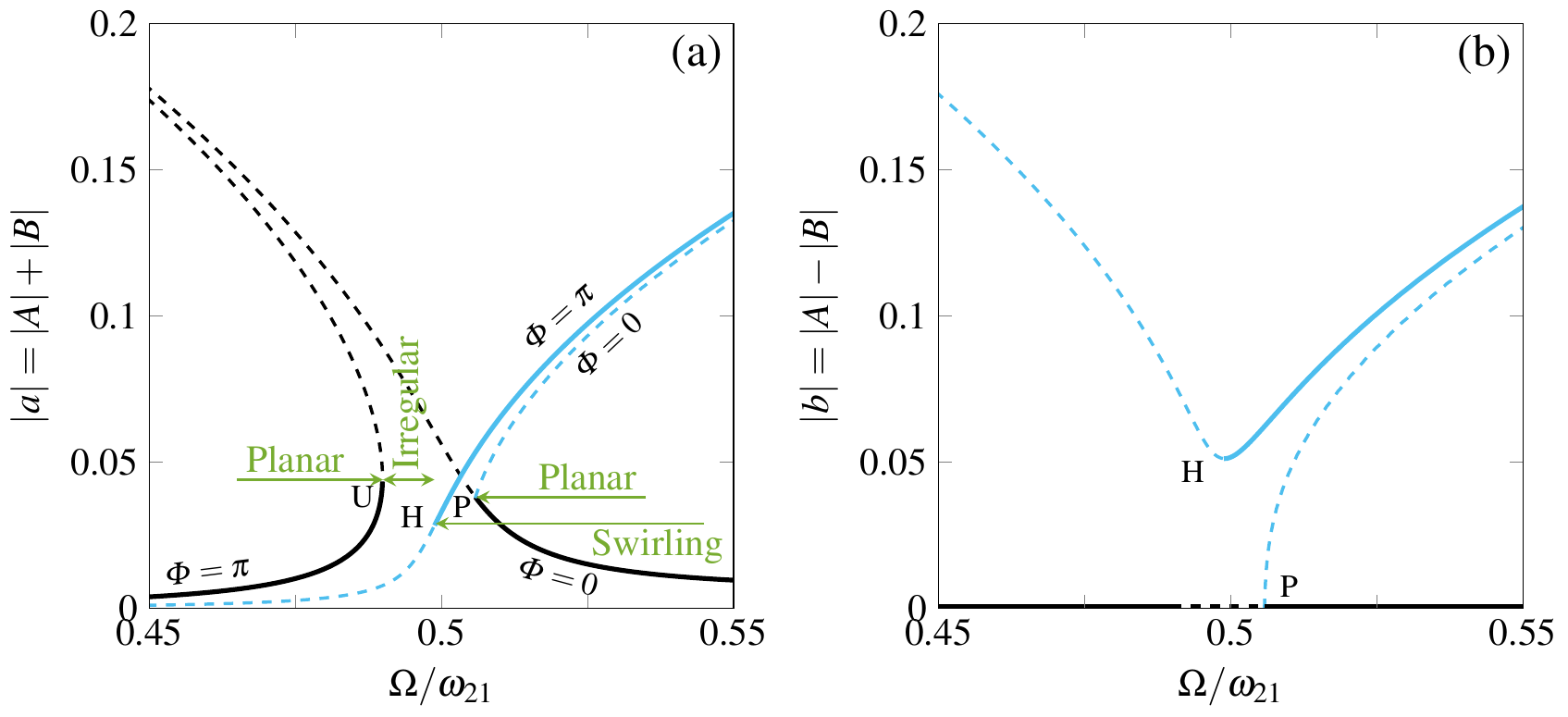}
\caption{Typical response curve for $a$ and $b$ for a fluid depth $H=1.5$ with longitudinal super-harmonic forcing of amplitude $a_x=0.2$. Panel (a) shows a projection of the three-dimensional \textcolor{black}{branch structure} $\left(\Omega/\omega_{21},|a|,|b|\right)$ in the $\left(\Omega/\omega_{21},|a|\right)$--plane, whereas panel (b) shows the same projection, but on the $\left(\Omega/\omega_{21},|b|\right)$--plane. \textcolor{black}{Black solid lines mark stable steady-state planar waves, whereas light blue solid lines indicate stable steady-state swirling waves. Dashed lines denote the corresponding unstable branches.} U: turning point. H: Hopf bifurcation. P: Poincaré bifurcation. For completeness, the phase values $\Phi_A=\Phi_B=\Phi=0$ or $\pi$ associated to each branch are reported in panel (a).}
\label{fig:Fig5} 
\end{figure}

As discussed in \cite{bongarzone2022amplitude} for rotary sloshing, although the quantitative dependence on the external control parameters, i.e. driving amplitude and frequency, is different with respect to the SC case, e.g. $f^2$ instead of $f$, system~\eqref{eq:AmpEqDCfinal_A}-\eqref{eq:AmpEqDCfinal_B} is essentially analogous to that given in~\eqref{eq:AmpEqSCfinalA}-\eqref{eq:AmpEqSCfinalB}. Indeed, equations~\eqref{eq:AmpEqDCfinal_A}-\eqref{eq:AmpEqDCfinal_B} contain four main contributions,
\begin{equation}
\label{eq:analogy_SC_DC}
\lambda \leftrightarrow \left(2\textcolor{black}{\lambda}-\frac{\chi_{_{DC}}}{4}f^2\right),\ \ \ \mu_{_{SC}}f\leftrightarrow \frac{\zeta_{_{DC}}\lambda+\mu_{_{DC}}}{4}f^2,\ \ \ \nu_{_{SC}}\leftrightarrow\nu_{_{DC}},\ \ \ \xi_{_{SC}}\leftrightarrow \xi_{_{DC}},
\end{equation}
\noindent corresponding respectively to a detuning term (forcing amplitude dependent), an additive (quadratic) forcing term (driving frequency dependent), the classic cubic restoring term and, lastly, the cubic term dictating the nonlinear interaction between the two counter-propagating traveling waves. For these reasons, figure~\ref{fig:Fig5} shows the nonlinear amplitude saturation for $|a|=|A|+|B|$ and $|b|=|A|-|B|$ which are reminiscent of those commented and displayed by \cite{faltinsen2016resonant} in their figure~7 with regard to harmonic system responses, although the phases associated to each super-harmonic branch are $\pi$-shifted with respect to the their harmonic analogous.\\
\begin{figure}
%\centering
\hspace*{1.75cm}\includegraphics[width=0.7\textwidth]{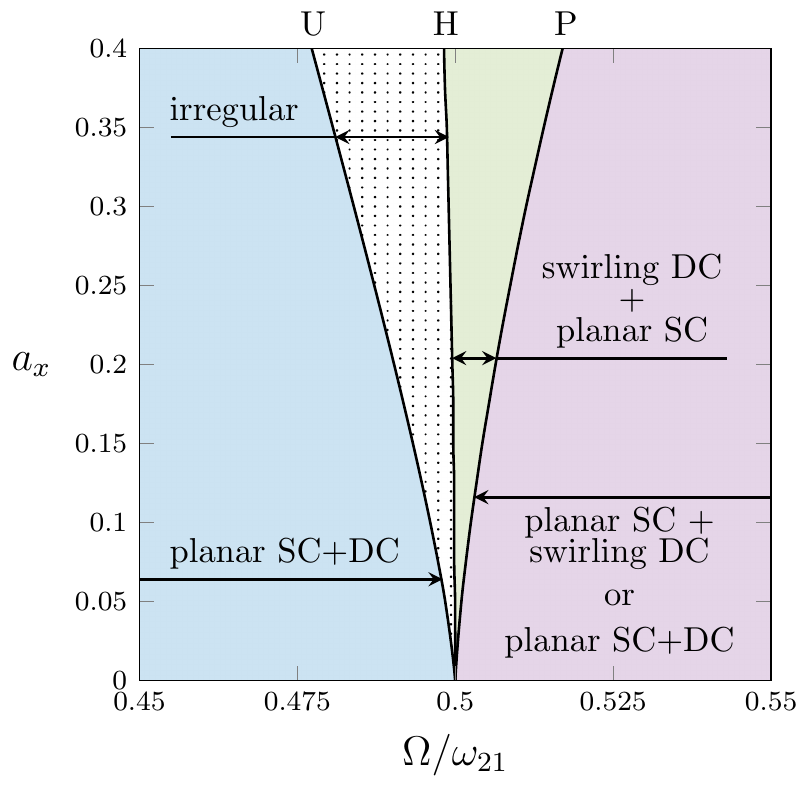}
\caption{Estimates of bounds, in the $\left(\Omega/\omega_{21},a_x\right)$-plane, between the frequency ranges where planar, irregular and swirling waves occur when the container undergoes a longitudinal and super-harmonic motion at a forcing frequency $\Omega\approx\omega_{21}/2$. In this range of frequency, the theory predicts the superposition of an unconditionally stable planar single-crest (SC) wave (\textcolor{black}{$m=\pm1$}) oscillating harmonically with the driving frequency and a super-harmonic double-crest (DC) dynamics (\textcolor{black}{$m=\pm2$}), which can manifest itself via planar, swirling or irregular wave motions. The stability boundaries (black solid lines) were computed for a fluid depth $H=1.5$, as in \cite{royon2007liquid}. The corresponding values of the normal form coefficients appearing in~\eqref{eq:AmpEqDCfinal_A}-\eqref{eq:AmpEqDCfinal_B} are given in table~\ref{tab:Tab3}.}
\label{fig:Fig6} 
\end{figure}
\indent A more detailed description of the bifurcation diagrams shown in figure~\ref{fig:Fig5}(a) and~(b) is given in \cite{faltinsen2016resonant}. Here we limit to note that the branching diagrams contain three bifurcation points, namely U (turning point), H (Hopf bifurcation) and P (Poincaré bifurcation), whose positions determine the frequency ranges where stable planar (standing), swirling or irregular waves are theoretically expected. By keeping track of the position of these three bifurcations points in the $\left(\Omega/\omega_{21},|a|\right)$-plane as the forcing amplitude, $a_x$, is varied, one can draw a super-harmonic stability chart in the $\left(\Omega/\omega_{21},a_x\right)$-plane similar to that of figure~\ref{fig:Fig3} for harmonic resonances and which is shown in figure~\ref{fig:Fig6}.\\
\indent The first striking difference with respect to the harmonic stability chart of figure~\ref{fig:Fig4} is the opposite curvature of the stability boundaries between the various super-harmonic regimes. As mentioned above, this is due to the quantitative dependence of the additive forcing term in system~\eqref{eq:AmpEqDCfinal_A}-\eqref{eq:AmpEqDCfinal_B} on the driving amplitude, which is here quadratic in $f$, thus leading to the square root in equations~\eqref{eq:planar_DC_sol} (planar DC) and~\eqref{eq:swirl_DC_sol_a} (swirling DC).

Furthermore, there is a substantial difference in terms of free surface patterns. As suggested by the form of the first order solution~\eqref{eq:DC_sol_eps1}, the leading order dynamics, governing the super-harmonic system response to longitudinal forcing, results from a superposition of a stable planar (or standing) single-crest (SC) wave, oscillating harmonically at a frequency $\Omega\approx\omega_{2n}/2$ and generated by \textcolor{black}{the two $m=\pm1$ counter-rotating traveling waves of equal amplitudes}, and a super-harmonic double-crest (DC) wave dynamics oscillating at a frequency of approximately $\omega_{2n}\approx2\Omega$ (period-halving). When the amplitudes of the two traveling waves with $m=\pm2$ are \textcolor{black}{equal}, i.e. $|A|=|B|$ (or $|b|=0$), the DC dynamics manifests itself via planar motion and the global solution takes the form of a planar wave (planar SC+DC, light blue shaded region in figure~\ref{fig:Fig6}). On the contrary, when $|A|\ne|B|\ne0$, one of the two $m=\pm2$ waves dominates over the other and a stable swirling motion, responsible of the system symmetry-breaking, is established. In this case, the total solution is given by the sum of a harmonic planar SC wave and a super-harmonic swirling DC wave (swirling DC+planar SC, green shaded region in figure~\ref{fig:Fig6}). The white-dotted region and the light red shaded regions in figure~\ref{fig:Fig6} correspond, respectively, to the super-harmonic irregular motion regime (see \S\ref{sec:Sec6} for further details) and to the multi-solution range where both types of motion are possible depending on the initial conditions, i.e. to the region of hysteresis.

\section{Experiments}\label{sec:Sec6}

In this section, we present our experimental set-up dedicated to the generation and characterization of sloshing waves under longitudinal super-harmonic forcing with driving (dimensionless) frequency $\Omega \approx \omega_{21}/2$. The bounds between the different regimes for the resulting super-harmonic wave are experimentally retrieved as a function of the driving amplitude and frequency, and compared to the theoretical estimates. Finally, we measure the wave amplitude saturation in the vicinity of the super-harmonic resonance, and compare it with the theoretical weakly nonlinear prediction~\eqref{eq:DC_sol_reconst}.

\subsection{Experimental set-up}

\begin{figure}
    \centering
    \includegraphics[width=0.6\textwidth]{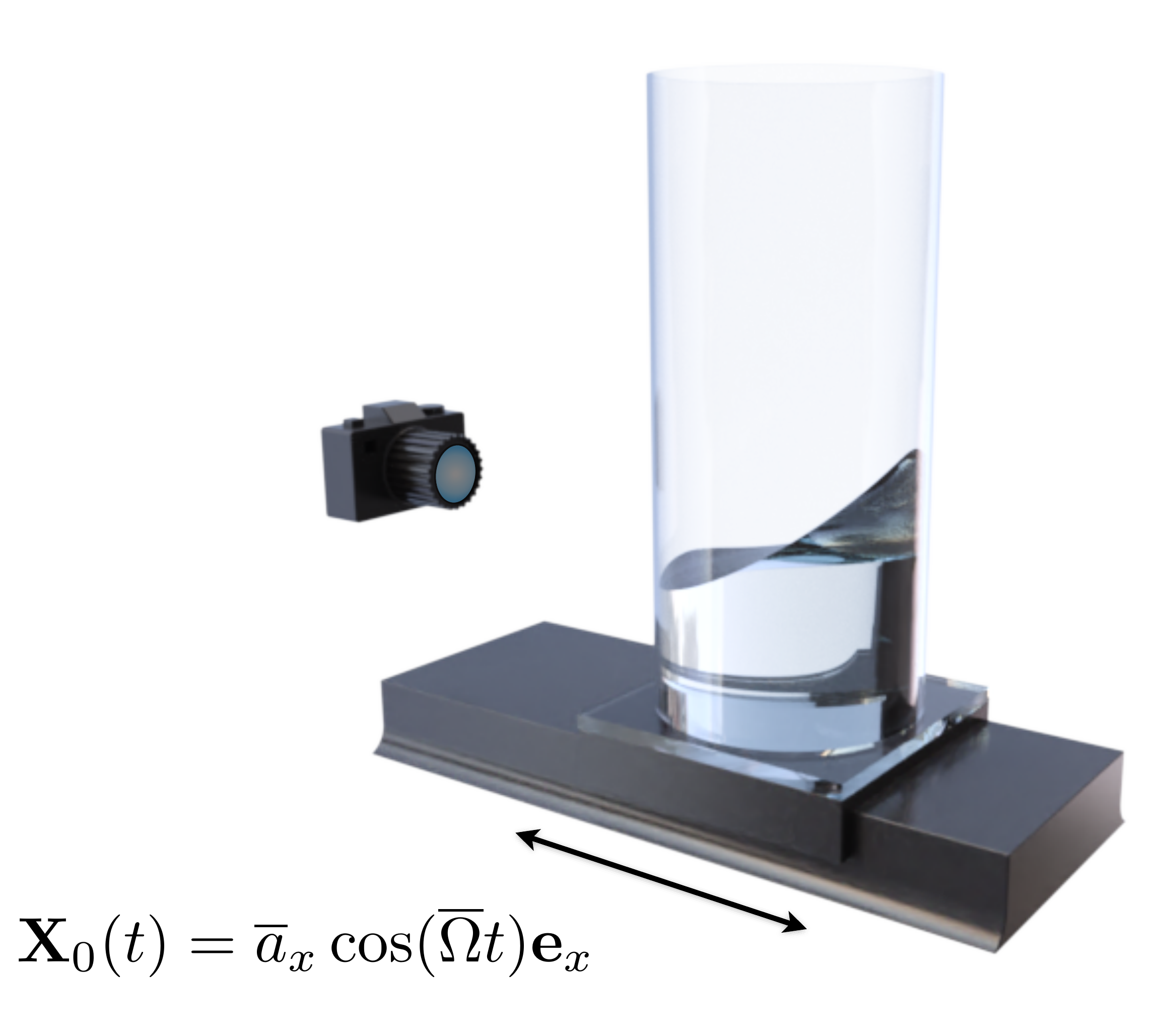}
    \caption{Experimental apparatus.}
    \label{fig:Fig_apparatus}
\end{figure}

The experimental set-up used to generate the sloshing waves in the cylindrical container and to observe the resulting free-surface motion is shown in figure \ref{fig:Fig_apparatus}. A Plexiglas cylindrical container of height 50 cm and inner diameter $D=2R=$ 17.2 cm, partially filled with a column of distilled water of height $h=$ 11 cm, is fixed on a single-axis linear motion actuator (AEROTECH PRO165LM). Sloshing waves are generated by imposing to the container a longitudinal sinusoidal forcing of angular frequency \textcolor{black}{$\bar{\Omega}$} and amplitude $\bar{a}_x$.   % précision du forçage en terme d'amplitude et de fréquence
\indent The motion of the fluid free-surface is recorded with a digital camera (NIKON D850) coupled with a Nikon 60mm f/2.8D lens and operated in slow motion mode, allowing for an acquisition frequency of 120 frames per second. The optical axis of the camera is aligned with the container motion axis. A LED panel (not depicted in Figure \ref{fig:Fig_apparatus}) placed behind the tank provides back illumination of the fluid free surface for a better optical contrast.  

The actuation of the moving stage as well as the camera triggering for movie recording are set and controlled via a home-made Labview program. In a typical experiment, the container undergoes a harmonic motion of fixed amplitude in the range 4 mm $\leq \bar{a}_x \leq$ 34 mm (i.e. $a_x = \bar{a}_x /R \in [0.05, 0.40]$), while a sweep in forcing frequency is implemented within the interval $\bar{\Omega}/ 2\pi \in$ [1.35 Hz, 1.58 Hz] corresponding to the dimensionless range $\Omega / \omega_{21} \in [0.45, 0.53]$. Each frequency step lasts 100 oscillation periods while the frequency increment between two consecutive steps is typically of 10 mHz. Along the sweeping, a movie is recorded for each $(\bar{a}_x, \bar{\Omega})$ set of parameters. To ensure that the steady-state amplitude regime is established at each step in the recorded free-surface dynamics, the camera is triggered only after a certain number of cycles, typically 50, see Appendix \ref{sec:AppD}.

\subsection{Analysis of the free-surface dynamics}

\subsubsection{Qualitative observations}

\begin{figure}
    \centering
    \includegraphics[width=0.95\textwidth]{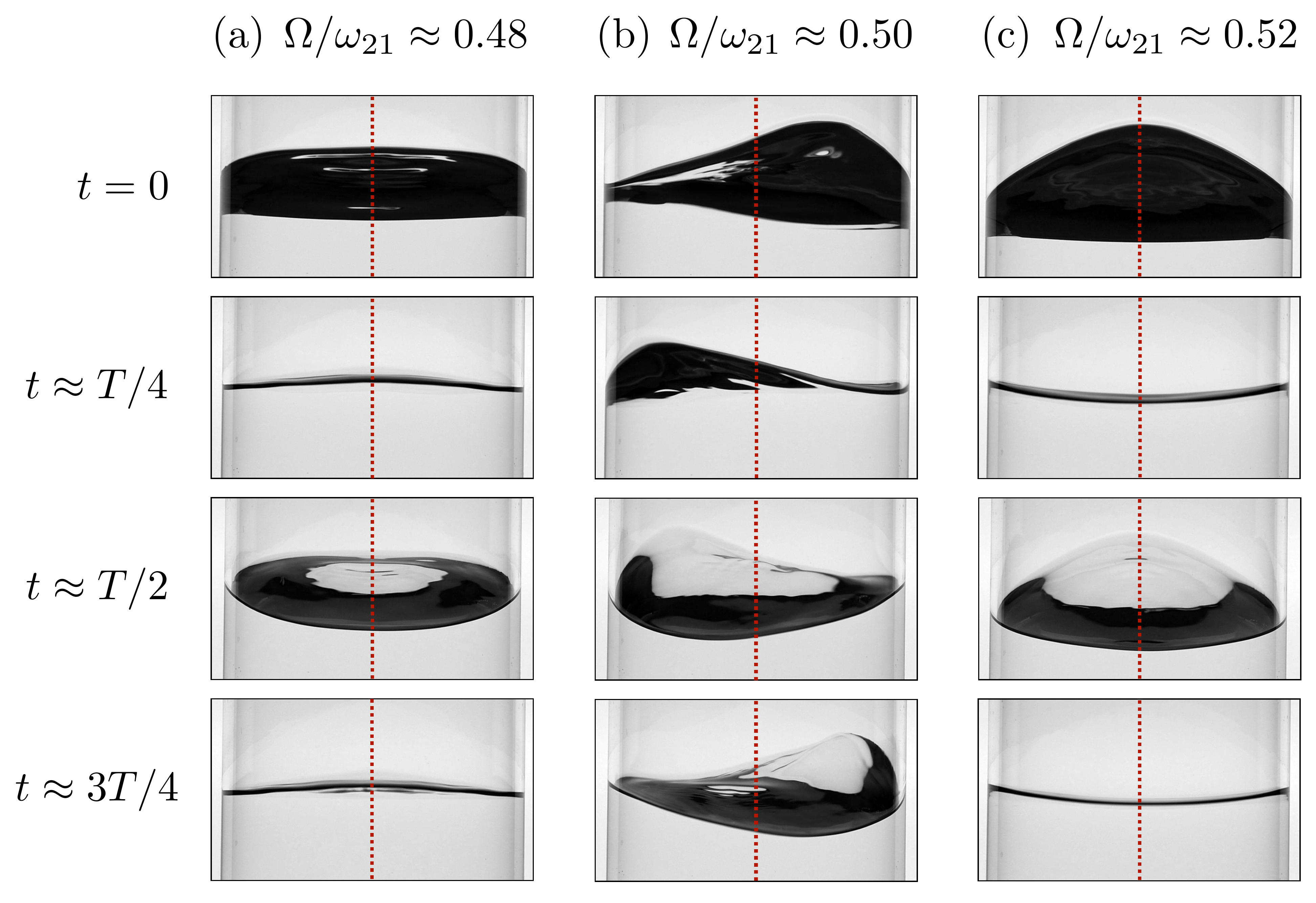}
    \caption{Images of the fluid free surface while the container is subjected to a longitudinal harmonic forcing of amplitude $a_x=\bar{a}_x/R \approx 0.23$ at various driving angular frequencies $\Omega$ close to $\omega_{21}/2$. The fluid free surface is observed in the direction aligned with the container motion. For each driving frequency (a)-(c), the time interval between two snapshots is about $T/4$, with $T=2 \pi / \Omega$ the corresponding oscillation period. On each snapshot, the vertical middle axis is represented by a red dotted line. For a forcing frequency $\Omega \approx 0.48 \omega_{21}$ (a) and $\Omega \approx 0.52  \omega_{21}$ (c) the free-surface image at each time $t$ is mirror-symmetric with respect to the middle vertical axis, signature of a planar motion, while the symmetry is broken for $\Omega \approx 0.50 \omega_{21}$ revealing a traveling swirling wave.}
    \label{fig:fig_snapshots}.
\end{figure}

While operating a sweep in forcing frequency at fixed forcing amplitude, we observe in the vicinity of the super-harmonic resonance three different kinds of motion, namely planar, irregular and swirling ones, whose occurrence depends on the forcing amplitude and frequency, see for instance the snapshots displayed on figure \ref{fig:fig_snapshots} or \textcolor{black} {the videos provided among the Supplementary Materials: (LINK).} 

For a given (and large enough) amplitude and starting from a frequency higher than a certain amplitude-dependent threshold $\Omega_P(a_x)$, the free surface responds to the longitudinal harmonic forcing \textcolor{black}{by displaying a planar dynamics} such as shown in figure \ref{fig:fig_snapshots}(c). When the critical frequency \textcolor{black}{$\Omega =\Omega_P(a_x)$} is reached, \textcolor{black}{the motion bifurcates to a swirling wave}, which propagates along the container wall with a stationary amplitude, see figure \ref{fig:fig_snapshots}(b). The wave can rotate either clockwise or anti-clockwise (both rotation directions were observed along the experiments). When the forcing frequency is further decreased below a critical frequency \textcolor{black}{$\Omega  = \Omega_H(a_x) \approx \omega_{21}/2 < \Omega_P(a_x)$}, the free surface exhibits an irregular dynamics, characterized by a switching between planar and swirling motion (not shown in figure \ref{fig:fig_snapshots}). For forcing frequencies lower than a certain threshold $\Omega < \Omega_U(a_x)$, the free surface motion stabilizes into a steady planar wave such as shown on figure \ref{fig:fig_snapshots}(a). 
%The snapshots displayed in figure \ref{fig:fig_snapshots} also provide some hints regarding the frequency content of the sloshing waves in the vicinity of the super-harmonic resonance $\Omega \approx \omega_{21}/2$. Indeed, it is clear than the free surface shape at any moment strongly differs here from the (flat) spatial structure of the singe-crest harmonic component figure \ref{fig:Fig_ModeDC}(a) and instead exhibits a saddle-like shape, which suggests the contribution of a resonating super-harmonic double-crest wave (figure \ref{fig:Fig_ModeDC}(b)) superposed to the harmonic, single-crest component. This contribution is easily detectable, as for instance in the planar regimes corresponding to $\Omega < \Omega_U(a_x)$ (figure \ref{fig:fig_snapshots}(a)) and $\Omega > \Omega_P(a_x)$ (figure \ref{fig:fig_snapshots}(c). In both cases, the harmonic single-crest component is in phase with the harmonic forcing (as suggested by the maximal front contact line elevation occurring 
%This is particularly obvious in the case of a swirling wave (figure \ref{fig:fig_snapshots}(b)) where the free surface clearly exhibits a saddle-like shape. 
\indent All together, these observations are qualitatively consistent with the outcomes of the weakly nonlinear analysis of Section \ref{sec:Sec5}, that predicts the existence of three different dynamical regimes -namely planar, irregular and swirling motion-, for a longitudinal forcing frequency \textcolor{black}in the vicinity of $\omega_{21}/2$. One of the main purposes of the present experimental investigation is to determine the amplitude-dependent frequency bounds of these different regimes and to compare them to our theoretical prediction of the positions of the bifurcation points U (turning point), H (Hopf bifurcation) and P (Poincaré bifurcation) (see figures \ref{fig:Fig5} and \ref{fig:Fig6}).

\subsubsection{Procedure}

Since the camera optical axis is aligned with the direction of the container motion, we note that a planar wave is characterized by its symmetry with respect to the vertical middle axis of the container image, whereas a swirling wave breaks this symmetry while traveling clockwise or anti-clockwise along the container walls, see figure \ref{fig:fig_snapshots}. 

We take benefit of these observations to build a more quantitative description of the free-surface dynamics, with the aim of identifying the various types of sloshing waves in the vicinity of the super-harmonic resonance. This can be done by exploiting the symmetry properties of the image of the free surface response with respect to the vertical middle axis of the container image, and by characterizing the regularity of these waves as a function of the forcing parameters, so as to identify the irregular regime. 
\begin{figure}
    \centering
    \includegraphics[width=0.95\textwidth]{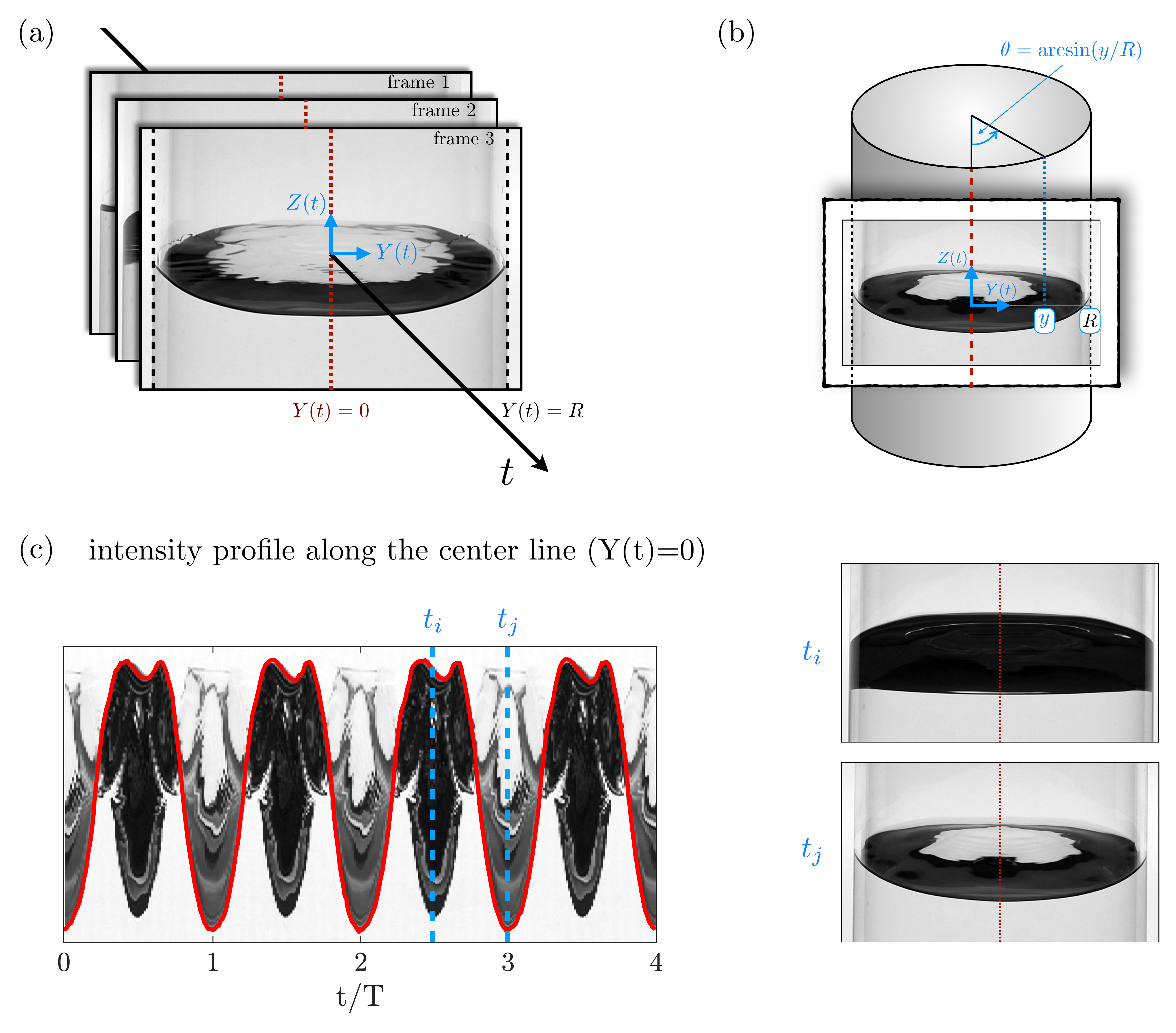}
    \caption{General procedure for the analysis of the free surface dynamics. (a) On each frame, the edges of the container are detected (black dotted lines) and the vertical $Z(t)$ axis is set as the middle line between these two edges, while the scale of the horizontal direction $Y(t)$ is fixed by the distance between both edges. (b) Schematic of the container illustrating the link between the Cartesian coordinate system $(Y(t), Z(t))$ attached to each frame, and the cylindrical coordinates in the referential frame of the container. (c) \textit{Left,} the intensity profile along a vertical line of coordinate $(Y(t)=y)$ with $y \in [-R, R]$ is then measured on each frame $t$ and plot as a function of time (here for $y=0$). The position of the front contact line at the azimuthal coordinate $\theta=\arcsin(y/R)=0$ as a function of time is highlighted in red. \textit{Right,} frames from which the intensity profiles at times $t_i$ and $t_j$ on the left-hand side image, along the line $(Y(t)=0)$ (represented by a red dotted line), are extracted. At $t_i$, the wave is climbing the front wall of the container (with respect to the camera position) whereas at $t_j$, it reaches the back of the tank.}
    \label{fig:fig_procedure}
\end{figure}
\indent To do so, the time evolution of the free surface dynamics is extracted from the movies along vertical directions that are mirror-symmetric with respect to the vertical middle axis of the container image. Comparing the resulting temporal signals with each other allows one to discriminate between planar and swirling motions and to study the wave regularity.  

The first step is to attach to each frame $t$ of a given movie, a Cartesian reference frame $(Y(t), Z(t))$, such that $Y(t)=0$ corresponds to the vertical middle axis of the container image, and that $Y(t)=R$ represents the right hand-side edge of the container image. To this end, the edges of the container are automatically detected in a dedicated Matlab program. The vertical $Z(t)$ axis \textcolor{black}{($Y(t)=0$)} on the frame corresponding to time $t$ is then set as the middle line between these two edges, while the distance between both edges sets the scale of the horizontal direction $Y$. Note that we neglect the 4 mm-thickness of the container wall. 

A direction $y \in [-R, R]$ is then chosen to extract from each frame corresponding to time $t_i$, the intensity profile $I_{t_i}(y)$ along the vertical line $Y(t_i)=y$. The resulting intensity profiles are then plotted as a function of time to build an image $I(y)$ composed as $I\left(y\right)=[I_{t_1}(y), I_{t_2}(y), ... ]$  such as displayed in figure \ref{fig:fig_procedure}(c). 
%In other words, the resulting image consists in an extraction from the movie of the plane $(Z(t), t)$ of coordinate $Y(t)=y$, where the system of coordinates $(Y(t), Z(t))$ is chosen on each frame so that $Z(t)$ represents the middle axis of the container image, and $Y(t)$ is scaled such that the vertical line $Y(t)=R$ represents the right hand-side edge of the container image. 

We note that at each time $t$, the intensity profile $I_{t}(y)$ contains the intersection of the front contact line image with the vertical axis $(Y(t)=y)$, that corresponds to the point of coordinates $(R, \theta, \eta(R, \theta, t))$ in the moving cylindrical frame of reference of the container, where $\theta=\arcsin(y/R)$ (see figure \ref{fig:fig_procedure}(b)). As a consequence, the final image $I(y)$ also contains the dynamics of the front contact line in the azimuthal direction $\theta$.

The resulting image $I(y)$ exhibits a periodic dark pattern that represents the free surface response to the harmonic forcing, see an example in figure \ref{fig:fig_procedure}(c) in which $y=0$. Indeed, on each frame of the movie, the free-surface appears as the darkest feature, so that the intensity profile along a given line $(Y(t)=y)$ actually represents the vertical extension of the free-surface at time $t$ along this direction, which is maximal whenever the sloshing wave reaches its maximal elevation $\max_t \eta( R, \theta, t)$ along the azimuthal direction $\theta= \arcsin(y/R)$ (in the front of the container with respect to the camera position, corresponding to $\theta \in ]-\pi/2, \pi/2[$) or along $\theta = \pi - \arcsin(y/R)$ (in the back of the container). Furthermore, when the contact line reaches its maximal elevation in the front of the container, the free-surface is imaged from below, so that it appears darker than when the maximal elevation is reached in the back, where the free-surface is imaged from above, see the snapshots in figure \ref{fig:fig_procedure}(c). These observations allow us to identify in the image $I(y)$ the position as a function of time of the front contact line $\eta ( R, \theta, t)$, with $\theta=\arcsin (y/R)$, as highlighted in red in figure \ref{fig:fig_procedure}(c). 

Note that this procedure does not give a quantitative access to the actual amplitude of the front contact line oscillations, since the intensity profiles $I_{t_i}(y)$ constituting the image $I(y)$ are simply juxtaposed with each other without rescaling the pixel width along the vertical direction. However, the position extracted from $I(y)$ of the image of the points of coordinates $\eta(R, \pm \theta, t)$ as a function of time still encloses the symmetry-properties of the free surface response, its regularity as well as its frequency content, which are the only quantities needed in order to identify the wave regimes.

\subsection{Regularity and frequency content of the free surface response}

\begin{figure}
    \centering
    \includegraphics[width=0.95\textwidth]{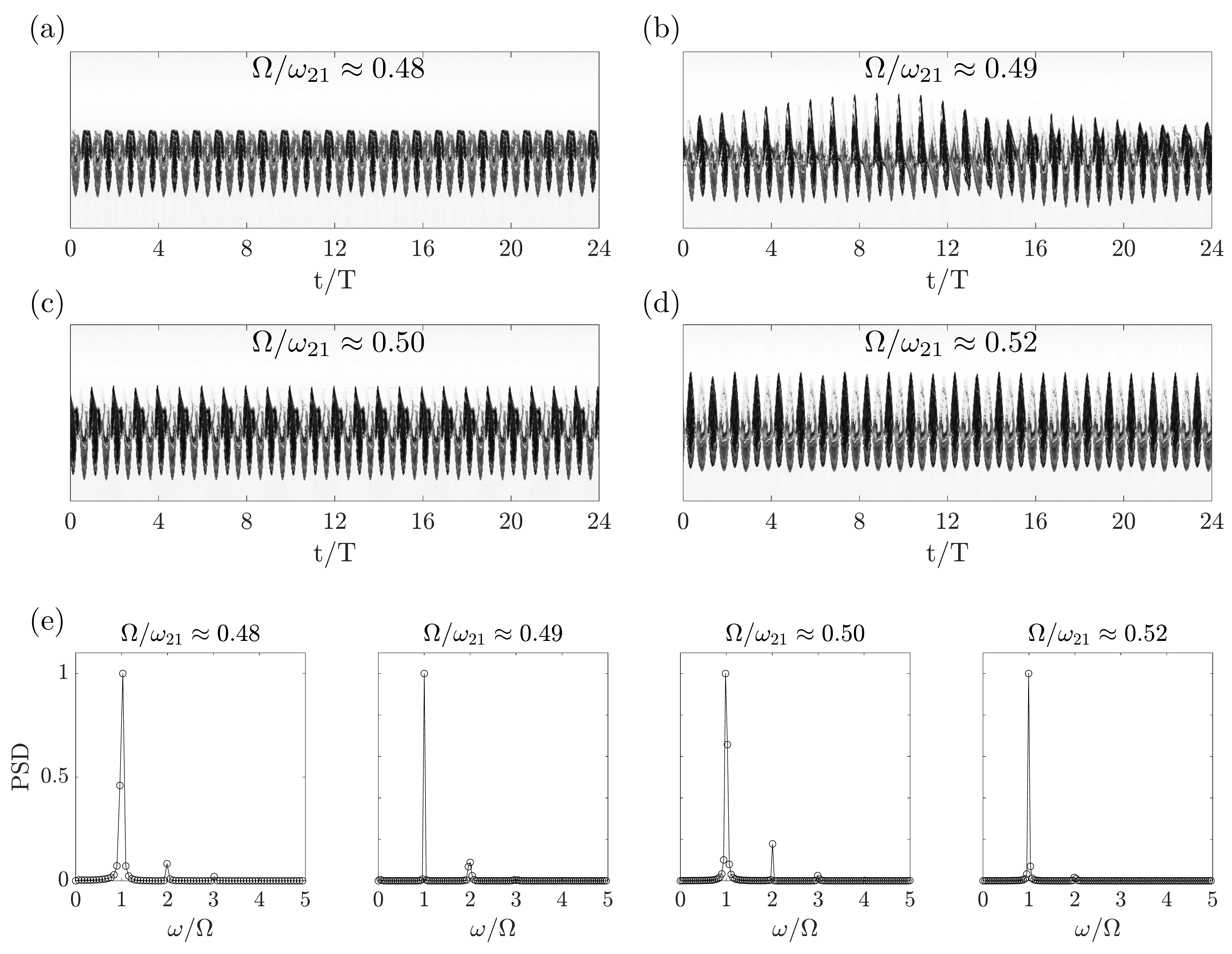}
    \caption{Panels (a) and (d): intensity profiles as a function of time along the vertical middle axis $(Y=0)$ -denoted $I(0)$- for various forcing frequencies $\Omega$ in the vicinity of the super-harmonic resonance $\Omega \approx \omega_{21}/2$ at same forcing amplitude $\bar{a}_x/R \approx 0.23$. In each case, the profile $I\left(0\right)$ is extracted from a movie whose recording has been started after about 50 oscillation cycles following each change in forcing frequency, thus ensuring that initial transients are filtered out (see also Appendix \ref{sec:AppD}). (e) Power spectral densities (PSD) -normalized by the maximal peak amplitude- corresponding to the front contact line dynamics as extracted from the profiles $I(0)$ displayed in panels (a) and (d).}
    \label{fig:fig_regularity}
\end{figure}
The resulting image $I(y)$ is then revealing of the free surface dynamics $\eta(r, \theta, t)$ and in particular of its dynamics at the front wall $\eta(r=R, \theta=\arcsin(y/R), t)$. Figure \ref{fig:fig_regularity}(a)-(d) displays $I(0)$ for various forcing frequencies close to the super-harmonic resonance, at same forcing amplitude. These images reveal that depending on the forcing frequency, the free surface oscillations (dark periodic pattern) can be either regular (a), (c) and (d) -i.e. the oscillations are enclosed into an envelope of constant amplitude- or irregular (b) with a temporal modulation of the amplitude envelope. Therefore, the profiles $I(0)$ allow to characterize the regularity of the sloshing wave, and in particular to identify the irregular regime. \textcolor{black}{The latter will be described in more details in Section \ref{sec:irregular_description}, but such details are not needed for the identification of the irregular regime bounds, for which the analysis of the regularity property of the $I(0)$-pattern is sufficient. Therefore in the following, we will focus on the regular planar and swirling motions, that cannot be unambiguously distinguished from each other on the basis of the profiles $I(0)$}. 

Figure \ref{fig:fig_regularity}(e) displays the (normalized) power spectral densities of the front contact line dynamics $\eta(R, \theta=0, t)$ extracted from the profiles $I(0)$ (a)-(d). It appears that in all cases, the energy of the sloshing wave is massively \textcolor{black}{distributed} to its first (harmonic) and second (super-harmonic) component, while the contribution of higher modes is fairly negligible. This incidentally implies that the symmetry properties of a regular wave are directly linked to the symmetry properties of these two first oscillation modes.

In other words, a planar dynamics should necessarily consist in the superposition of two planar waves: a planar single-crest (SC) wave harmonically oscillating with the driving frequency $\Omega$ and one super-harmonic planar double-crest (DC) wave oscillating at $\omega= 2 \Omega \approx \omega_{21}$. On the other hand, a swirling dynamics must contain at least one symmetry-breaking (swirling) component that, as predicted by the present weakly nonlinear analysis, should correspond to the super-harmonic $\omega_{21}$ component. 

\subsection{Symmetry properties of the regular regimes: planar versus swirling waves}

We now focus on the regular regimes, namely the steady planar and swirling motions. As stated before, the profiles $I(0)$ cannot discriminate between a planar and a swirling dynamics and instead only contain information on their regularity and their frequency content. To distinguish a planar from a swirling motion, we then compare the profiles along two ($Y \neq 0$)-directions that are symmetric with respect to the vertical middle axis of the container image. 
\begin{figure}
    \centering
    \includegraphics[width=0.8\textwidth]{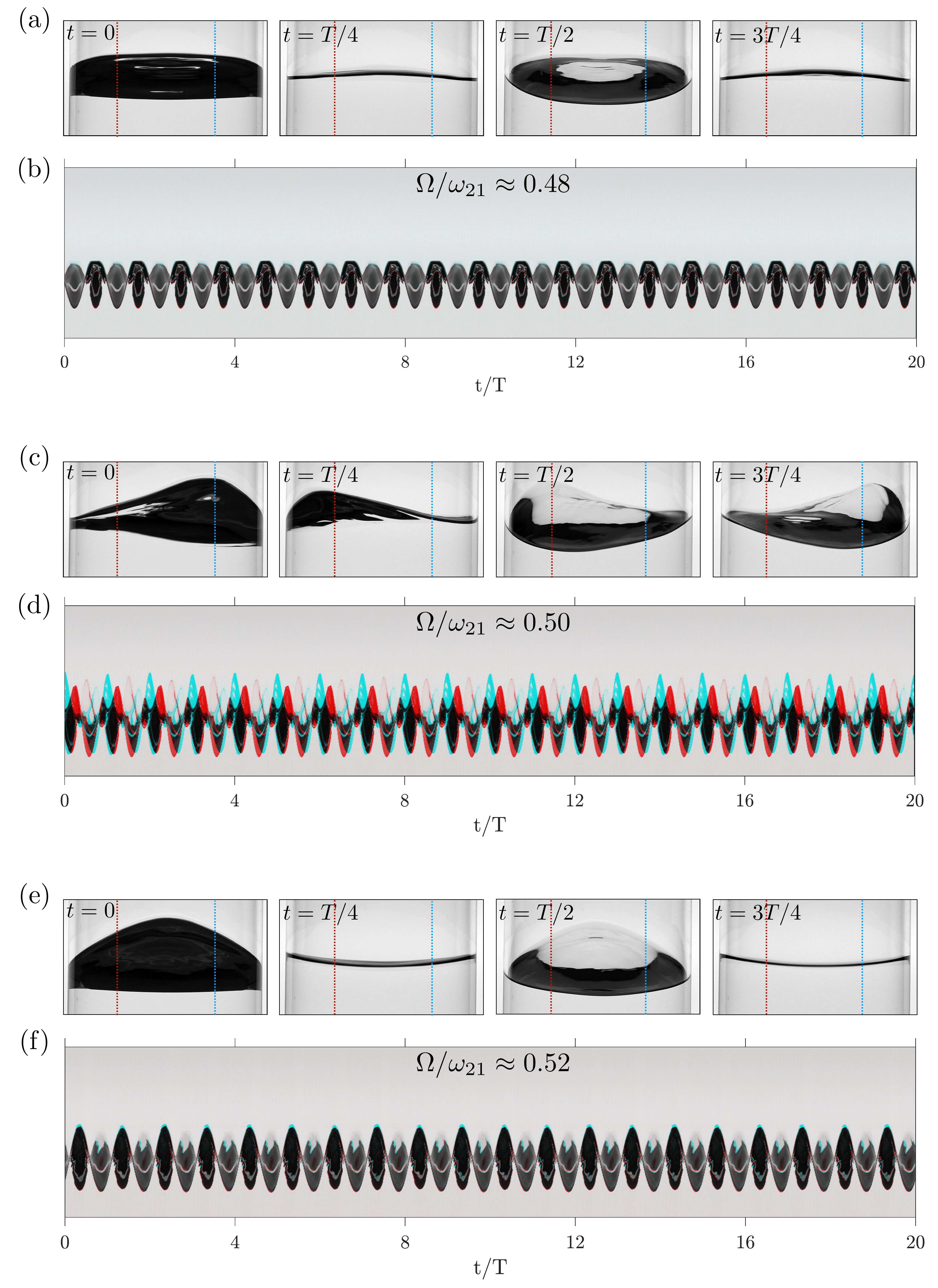}
    \caption{Symmetry properties of the stationary waves. (a), (c) and (e): images of the fluid free surface while the container is subjected to a longitudinal harmonic forcing of amplitude $\bar{a}_x/R \approx 0.23$ at various driving angular frequencies $\Omega$ close to $\omega_{21}/2$: (a) $\Omega \approx 0.48 \, \omega_{21}$, (c) $\Omega \approx 0.50\,  \omega_{21}$ and (e) $\Omega \approx 0.52 \, \omega_{21}$ (same forcing parameters as in figure \ref{fig:fig_regularity}(a), (c) and (d)). For each driving frequency (a, c, e), the time interval between two snapshots is about $T/4$, with $T=2 \pi / \Omega$ the corresponding oscillation period. On each snapshot, the vertical axes $(Y=R/2)$ and $(Y=-R/2)$ are represented by a blue and red dotted line, respectively. For a forcing frequency $\Omega=0.48 \omega_{21}$ (a) and $\Omega=0.52 \omega_{21}$ (e) the free-surface image at each time $t$ is mirror-symmetric with respect to the middle vertical axis, while the symmetry is broken for $\Omega = 0.50 \, \omega_{21}$. (b), (d) and (f): superposition of the intensity profiles as a function of time along the vertical axis $(Y=R/2)$ and $(Y=-R/2)$ -denoted $I(R/2)$ (in blue) and $I(-R/2)$ (in red) respectively-, for the same forcing parameters as in figure \ref{fig:fig_regularity}(a), (c), and (e). The gray regions show where $I(R/2)$ and $I(-R/2)$ have the same intensities.}
    \label{fig:fig_regular_regimes}
\end{figure}
\textcolor{black}{Figure \ref{fig:fig_regular_regimes}(b), (d) and (f) show composite images, each produced using the Matlab function \textit{imshowpair} applied to the pair \textcolor{black}{$I(R/2)$} and \textcolor{black}{$I(-R/2)$}, for three different forcing frequencies that both result in a regular motion (same forcing parameters as in figure \ref{fig:fig_regularity}(a), (c), and (d)). Briefly, \textit{imshowpair}$(A,B)$ creates from a pair of grayscale images $A$ and $B$, a RGB image where each pixel is represented by a RGB triplet, the R-intensity being the intensity of the corresponding pixel in $A$, and the G- and B-intensities being equal to the intensity of the corresponding pixel in $B$. A pixel where $A$ and $B$ have the same intensity will be represented by a RGB triplet of the forme $[a, a, a]$, where $a \in [0, 255]$, i.e. will appear as gray. On the contrary, if this pixel has a much larger intensity on $A$ (resp. on $B$) than it has on $B$ (resp. on $A$), it will appear in red (resp. in cyan) on the resulting composite image. The composite images displayed in Figure \ref{fig:fig_regular_regimes}(b), (d) and (f) thus highlight in each case the differences between $I(R/2)$ and $I(-R/2)$. } They are then a direct signature of the symmetry of the free-surface dynamics with respect to the vertical middle axis $(Y=0)$, and reveal two different kinds of motion, (i) a planar motion, for which $I(R/2)$ and $I(-R/2)$ perfectly overlap with each other due to the  mirror-symmetry of the wave, and (ii) a rotary motion, characterized by a symmetry-breaking between the right and left hand-side free-surface dynamics: the maximum of the wave along $\theta=\arcsin(1/2)=\pi/6$ is indeed phase-shifted with respect to the maximum of the wave along $\theta=-\pi/6$, thus revealing a traveling wave propagating along the wall of the container. 

\begin{figure}
    \centering
    \includegraphics[width=0.90\textwidth]{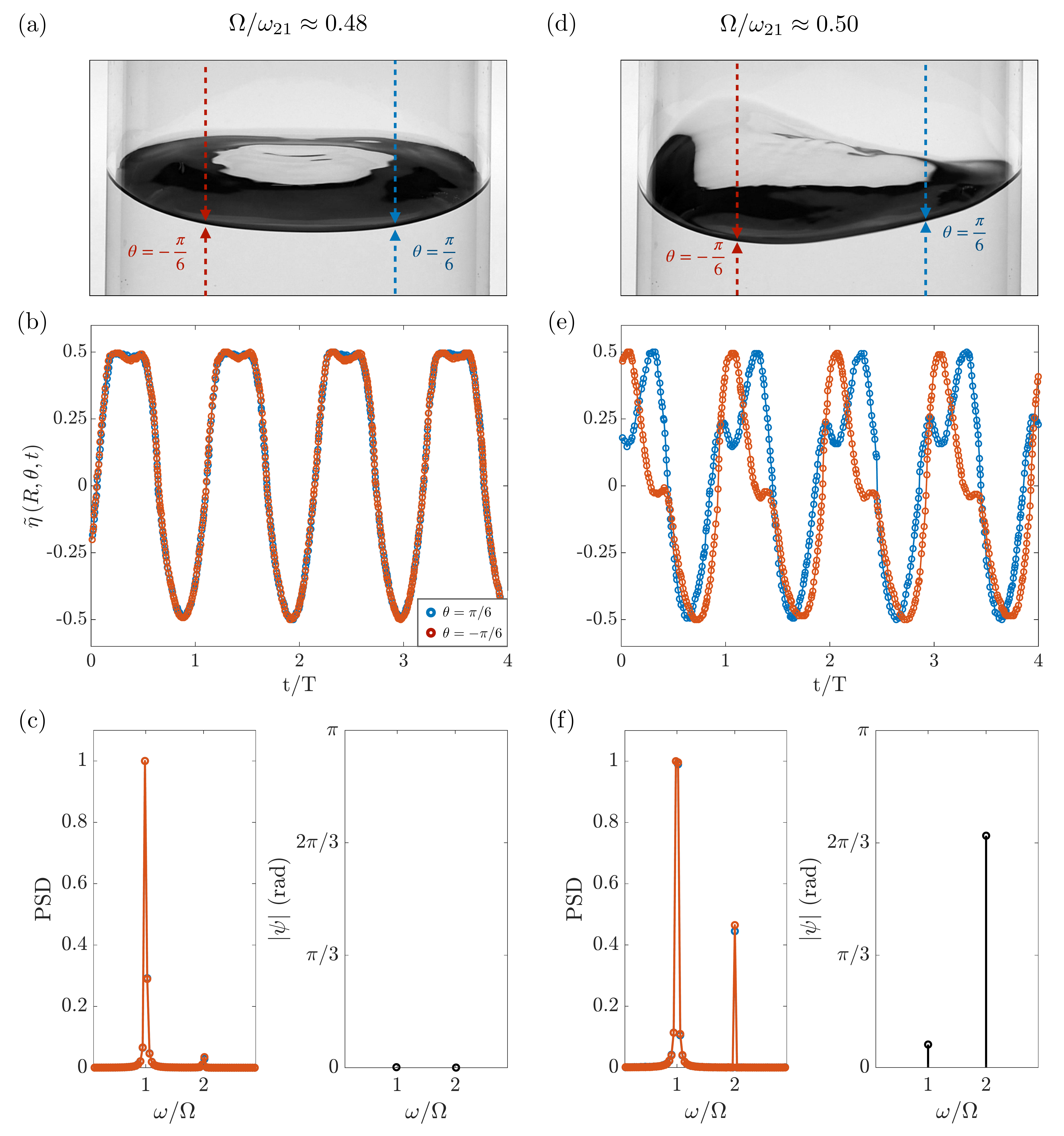}
    \caption{Analysis of the steady-state free-surface dynamics under an harmonic forcing of amplitude $a_x=0.23$ and frequency (a)-(c) $\Omega /\omega_{21} \approx 0.48$ and (d)-(f) $\Omega / \omega_{21} \approx 0.50$. (a) and (d) Image of the free surface with vertical lines intersecting the image of the front contact line at the points of coordinates $(R, \pi/6, \eta(R, \pi/6, t))$ (blue arrows) and $(R, -\pi/6, \eta(R, -\pi/6, t))$ (red arrows) in the moving reference frame of the container. (b) and (e) Normalized elevation of the front contact line $\tilde{\eta}(R, \pi/6, t)$ (blue dots) and $\tilde{\eta}(R, -\pi/6, t)$ (red dots) extracted from the corresponding profiles $I(R/2)$ and $I(-R/2)$ (not shown here). The $\tilde{\eta}$-functions are defined according to $\tilde{\eta}(R, \theta, t)= \left(\eta(R, \theta, t)-\sigma \right) / \delta$, where $\sigma=\left(\min_t(\eta(R, \theta, t)+\max_t(\eta(R, \theta, t) \right)/2$ and $\delta=\max_t(\eta(R, \theta, t))-\min_t(\eta(R, \theta, t))$. (c) and (f) \textit{Left}, Power spectral densities of $\tilde{\eta}(R, \pi/6, t)$ (blue dots) and of $\tilde{\eta}(R, -\pi/6, t)$ (red dots). \textcolor{black}{\textit{Right}, Absolute value of the phase shift between the components of $\tilde{\eta}(R, \pi/6, t)$ and of $\tilde{\eta}(R, -\pi/6, t)$ oscillating at the frequencies corresponding to the first peak ($\omega = \Omega$) and to the second peak ($\omega = 2 \Omega$) of the power spectra.}}
    \label{fig:fig_planar_vs_swirling}
\end{figure}

To determine which $\omega$-component is responsible for the symmetry-breaking induced by the swirling motion, we extract from $I(R/2)$ and $I(-R/2)$ the position of the front contact line as a function of time $\eta(R, \theta, t)$ where $\theta = \pm \pi/6$, see figure \ref{fig:fig_planar_vs_swirling}(b), (e). This makes possible to compute the power spectrum of both signals, \textcolor{black}{as well as the phase difference between the phase angle of their components that oscillate at the frequencies corresponding to their spectrum's first and second peaks (see figure \ref{fig:fig_planar_vs_swirling}(c-f))}. 
A planar wave oscillating at a frequency $\omega$ is then characterized by the $\omega$-components of $\eta (R, \pi/6, t)$ and of $\eta (R, -\pi/6, t)$ being in phase with each other, while a swirling wave is characterized by a $m \pi /3$-phase shift between the $\omega$-components of these signals, where $m$ denotes the azimuthal wavenumber of the swirling wave ($m=1$ for an harmonically oscillating single-crest wave, $m=2$ for a super-harmonic double-crest wave). 

The Fourier analysis of the signals $\eta(R, \pi/6, t)$ and $\eta(R, -\pi/6, t)$ reveals that for forcing frequencies $\Omega$ close to $\omega_{21}/2$, the free surface motion mostly results from the combination of a single-crest wave harmonically oscillating at the forcing frequency $\Omega \approx \omega_{21}/2$, and of a super-harmonic double-crest wave oscillating at a frequency $2 \Omega \approx \omega_{21}$. \textcolor{black}{From figure \ref{fig:fig_planar_vs_swirling}(c) and (f), it is clear that the single-crest wave is a planar wave for both planar (figure \ref{fig:fig_planar_vs_swirling}(c)) and swirling (figure \ref{fig:fig_planar_vs_swirling}(f)) dynamics, as revealed by the vanishing phase-shift between the harmonic components of $\eta(R, \pi/6, t)$ and $\eta(R, -\pi/6, t)$ in both cases. On the other hand, the phase shift between the super-harmonic components is zero in the case of the planar dynamics, and close to $2 \pi/3$ in the case of the swirling dynamics, signature of a double-crest swirling wave. }
\textcolor{black}{These observations are general to the whole range of forcing frequencies and amplitudes investigated along this study: in the vicinity of the super-harmonic resonance, the single-crest wave is always a planar wave, as revealed by the vanishing phase-shift between the harmonic components of $\eta(R, \pi/6, t)$ and $\eta(R, -\pi/6, t)$ for both planar and swirling dynamics (this is also true in the case of the irregular regime, see later Section \ref{sec:irregular_description}). In the case of a regular dynamics, the double-crest wave is either a planar (for vanishing phase-shift between the corresponding components) or a swirling wave (characterized by a $2 \pi/3$ phase-shift between the $\omega_{21}$-component of the right and left hand-side signals), depending on the exact ratio between $\Omega$ and $\omega_{21}$, as well as on the forcing amplitude $\bar{a}_x/R$. }

\subsection{Experimental estimate of regime bounds}\label{subsec:Exp_est_bounds}

From the above analysis, it appears that consistently with the predictions provided by our theoretical weakly nonlinear analysis, the sloshing waves resulting from the longitudinal super-harmonic forcing of the container at a frequency $\Omega \approx \omega_{21}$, consist in the superposition of \textcolor{black}{a planar single-crest wave}, harmonically oscillating with the forcing at $\omega =\Omega$, and of a double-crest wave, than can exhibit either a planar, irregular or swirling dynamics. 

Having identified the three different regimes for the free-surface dynamics in the vicinity of the super-harmonic resonance, we can now experimentally determine their stability regions in the $(\Omega/\omega_{21}, a_x)$ space. To do so, we fix the forcing amplitude while operating a frequency sweep from high to low frequencies, within the range $\Omega/\omega_{21} \in [0.45, 0.53]$, by frequency decrements of 10 mHz. Note that a downward frequency sweep ensures to recover the stability bound between the super-harmonic planar and swirling regimes, as the transition in this direction occurs exactly at the threshold frequency $\Omega_P(a_x)$ below which the super-harmonic planar motion becomes unstable. \textcolor{black}{On the contrary, since the super-harmonic swirling wave is still stable for frequencies larger than $\Omega_P(a_x)$ (hysteresis), an upward frequency sweep will maintain the system's response on the swirling branch, thus it is not suitable to experimentally detect the bifurcation point P.} 
\begin{figure}
\centering
\includegraphics[width=0.8\textwidth]{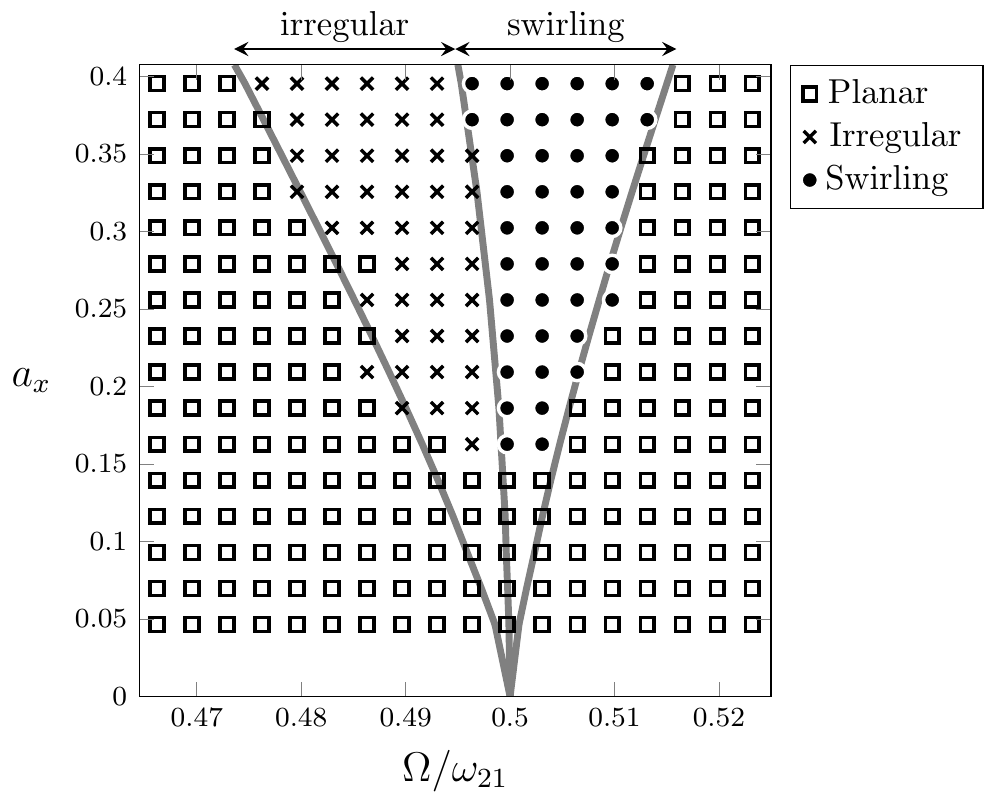}
\caption{Estimates of regime bounds in the $\left(\Omega/\omega_{21},a_x\right)$-plane for a container of diameter $D=0.172\,\text{m}$, filled to a depth $H=1.3$, driven longitudinally and super-harmonically at a frequency $\Omega\approx\omega_{21}/2$: comparison between the theoretical predictions (solid lines) and experimental measurements (markers). Gray thick solid lines: present theoretical predictions. Black empty squares: super-harmonic planar motion. Black crosses: irregular regime. Black filled circles: super-harmonic swirling motion.}
\label{fig:Fig_stability_regions} 
\end{figure}
%the system's response is hysteretic, causing an upward frequency sweep to maintain the swirling wave under forcing frequencies beyond the threshold, up to another jump-down frequency $\Omega > \Omega_P(a_x)$ that cannot be predicted by our inviscid analysis. That makes an upward frequency sweep unsuitable to experimentally determine the threshold frequency $\Omega_P(a_x)$.} 
The downward frequency sweep also enables one to detect the bounds that separate the irregular regime from steady planar ($\Omega = \Omega_U(a_x)$) and swirling motions ($\Omega = \Omega_H(a_x)$).

This procedure is applied for various forcing amplitudes $a_x \in [0.05, 0.4]$, enabling us to build the stability regions diagram displayed on figure \ref{fig:Fig_stability_regions}. All together, the experimental measurements are in very good quantitative agreement with the theoretical regime bounds for $a_x > 0.15$, below which the super-harmonic irregular and swirling regimes appear to be suppressed by dissipative mechanisms, e.g. viscous dissipation occurring in the fluid bulk, sidewall and free surface boundary layers \citep{Case1957,Miles67,raynovskyy2020sloshing,bongarzone2022sub} as well as in the neighbourhood of the moving contact line \citep{Keulegan59,Dussan79,Hocking87,Cocciaro93,Viola2018b}. Note that viscous dissipation is not accounted for in our theoretical asymptotic analysis based on an inviscid model.

\subsection{Irregular regime}\label{sec:irregular_description}

%\citep{chen2023mechanism}

\begin{figure}
    \centering
    \includegraphics[width=0.95\textwidth]{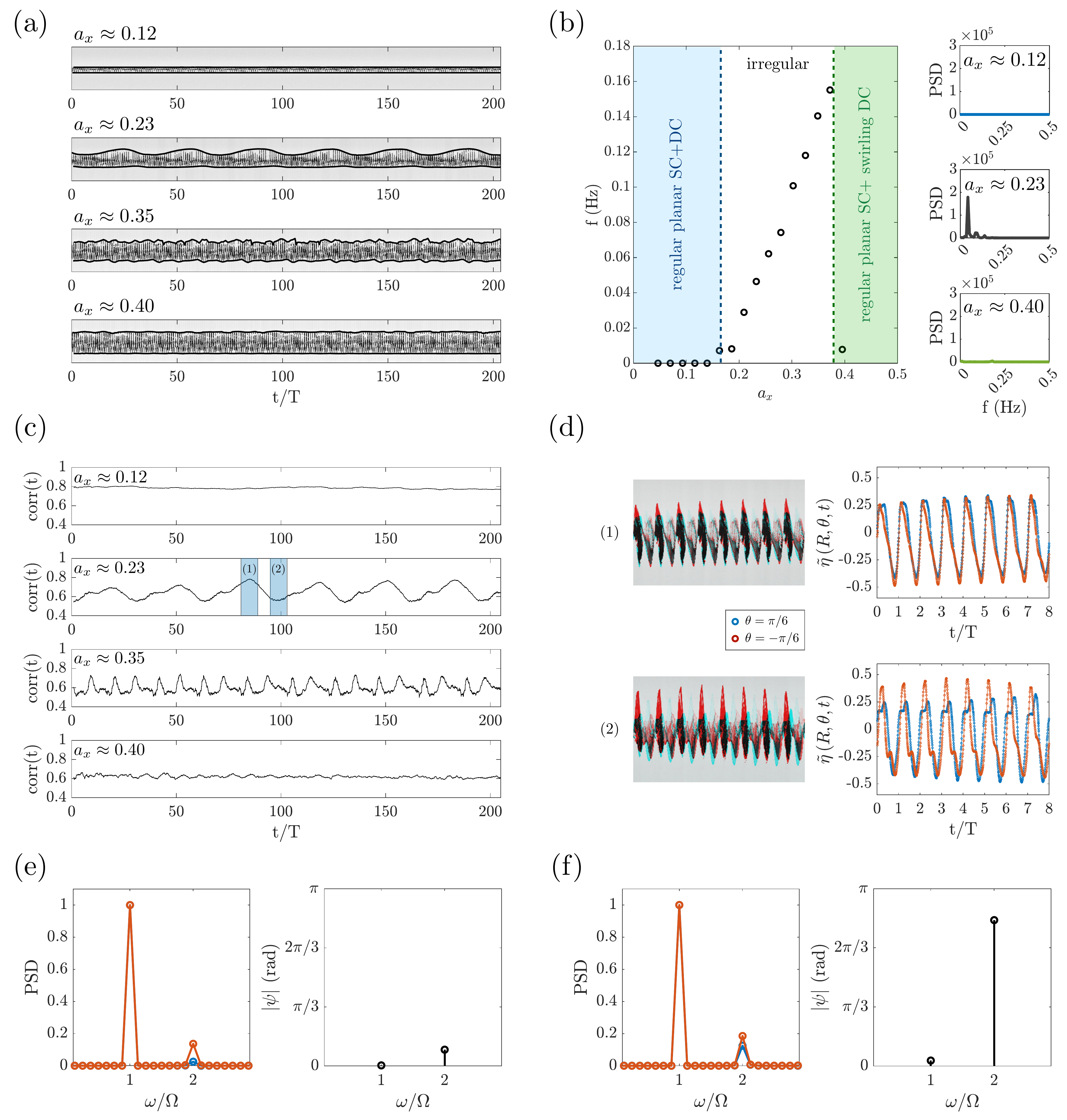}
    \caption{(a) Intensity profiles $I(0)$ for various forcing amplitudes and same forcing frequency $\Omega \approx 0.496 \omega_{21}$. The oscillations of the free surface are \textcolor{black}{enclosed into an envelope}, plotted in black on top of the images. (b) \textit{Right,} Frequency of the main peak in the power spectrum of the envelope as a function of the forcing amplitude $a_x$, for same forcing frequency $\Omega \approx 0.496 \omega_{21}$ . \textcolor{black}{When the envelope is a straight line (as it is the case here for $a_x < 0.15$, which corresponds to a regular planar dynamics of the free surface), the power spectrum is flat and we set the corresponding frequency equal to zero. For $a_x \approx 0.40$, the power spectrum of the envelope is dominated by a low amplitude and small frequency noise, causing a brutal decrease of the ``burst'' frequency, thus indicating a transition from irregular to regular swirling motion.}  \textit{Left,} Power spectra of the envelope for $a_x\approx 0.12$, $a_x \approx 0.23$ and $a_x \approx 0.40$. (c) Correlation between $I(R/2)$ and $I(-R/2)$ as a function of time, for the same set of forcing parameters as in (a). (d) \textit{Left,} Superposition of $I(R/2)$ (blue) and $I(-R/2)$ (red) and \textit{right,} position of the front contact line $\eta( R, \pm \pi/6, t)$ as a function of time extracted from $I(-R/2)$ (red curve) and from $I(R/2)$ (blue curve). The signals presented in (d) are taken from the full signals used to compute their correlation in (c) for $a_x \approx 0.23$, over the time ranges highlighted in blue and denoted as (1) (maximum of correlation) and as (2) (minimum of correlation). (e)-(f) Normalized power spectra of $\eta(R, -\pi/6, t)$ (red curve) and of $\eta(R, \pi/6, t)$ (blue curve), and absolute value of the phase-shift between their harmonic and super-harmonic components, where the dynamics of $\eta(R, \pm \pi/6, t)$ is considered over (e): time range (1) and (f): time range (2).}
    \label{fig:fig_irregular}
\end{figure}

In this section, we provide a more thorough description of the irregular regime. When fixing the forcing frequency slightly below $\omega_{21}/2$ and progressively increasing the forcing amplitude, the free-surface response is first very regular and displays a planar dynamics for low enough forcing amplitudes. Above a threshold amplitude, the dynamics becomes irregular and at large enough amplitudes, the response is again regular, but consists in a swirling motion. Figure \ref{fig:fig_irregular}(a) displays the free surface response along the vertical middle axis $Y=0$ for increasing forcing amplitudes at a fixed forcing frequency $\Omega \approx 0.496 \omega_{21}$. The regular regimes (top and bottom panels) are characterized by a constant amplitude of the free surface oscillations. In contrast, the oscillations of the free surface for intermediate forcing amplitudes (second and third panels) are enclosed into a quasi-periodic envelope, whose frequency linearly increases with the forcing amplitude (see figure \ref{fig:fig_irregular}(b)). This is very reminiscent of the observations by \cite{royon2007liquid} of the irregular regime present in the vicinity of the harmonic resonance under longitudinal forcing. 

To gain more insight on this irregular dynamics, we compute at each time $t_i$ the spatial correlation between $I_{t_i}(R/2)$ and $I_{t_i}(-R/2)$, which we refer to as corr$(t_i)$%, \textcolor{black}{and whose expression is given by:}

\begin{equation}
\text{corr}(t_i)=\frac{\sum_n \left(I_{t_i, n}(R/2)-\overline{I}_{t_i}(R/2)\right)\left(I_{t_i, n}(-R/2)-\overline{I}_{t_i}(-R/2)\right)}{\sqrt{\sum_n \left(I_{t_i, n}(R/2)-\overline{I}_{t_i}(R/2)\right)^2\sum_n \left(I_{t_i, n}(-R/2)-\overline{I}_{t_i}(-R/2)\right)^2}},
\end{equation}
\textcolor{black}{where $n \in [1, N]$, with $N$ the number of pixels in the vertical direction, and $\overline{I}_{t_i}(y)$ represents the mean of the N-element vector $I_{t_i}(y)$. }
A high and constant correlation is a signature of a steady planar motion, while a low but still constant correlation is characteristic of the steady swirling regime. At intermediary forcing amplitudes -i.e. in the irregular regime- the correlation is a quasi-periodic function of time, with the same quasi-period as the envelope, see figure \ref{fig:fig_irregular}(c). 

A comparison between $I(R/2)$ and $I(-R/2)$ on time ranges corresponding to maximum and minimum of the correlation function reveals that in the time interval where the signals are highly correlated, the motion is planar-like (although irregular), while in the time range where they are poorly correlated, the maxima of the right and left hand-side signals are phase-shifted with respect to each other, thus reflecting the presence of a swirling wave, see figure \ref{fig:fig_irregular}(d). 

This is further confirmed by the power spectra of the front contact line dynamics along the azimuthal directions $\theta=\pm \pi/6$, extracted from $I(R/2)$ and $I(-R/2)$, on time ranges where these signals are highly correlated and where they are poorly correlated, see figure \ref{fig:fig_irregular}(e-f). In both cases, the sloshing wave contains a planar single-crest wave, as revealed by the vanishing phase-shift between the harmonic components of $\eta(R, \pi/6, t)$ and $\eta(R, -\pi/6, t)$. The wave also contains a super-harmonic component, that is responsible for the switching between a planar-like motion (vanishing phase-shift between the $\omega_{21}$-components of the $\eta(R, \pi/6, t)$ and $\eta(R, -\pi/6, t)$ signals, \textcolor{black}{figure \ref{fig:fig_irregular}(e)}) and a swirling dynamics (rotating, symmetry-breaking wave that is super-harmonically oscillating at $\omega \approx \omega_{21}$, \textcolor{black}{figure \ref{fig:fig_irregular}(f)}). 

This is again very similar to the features of the irregular regime in the vicinity of the harmonic resonance described by \cite{royon2007liquid} that relate the ``bursts'' in the free-surface oscillation amplitude to the quasi-periodic occurrence of a swirling wave. However in the case of super-harmonic resonance, the irregular regime consists here in the superposition of a stable planar single-crest wave and of a super-harmonic double-crest dynamics. The latter is responsible for the irregularity of the total dynamics, by quasi-periodically switching between super-harmonic planar and swirling motion.

\subsection{Wave amplitude saturation: theoretical predictions versus experiments}

\begin{figure}
\centering
\includegraphics[width=1\textwidth]{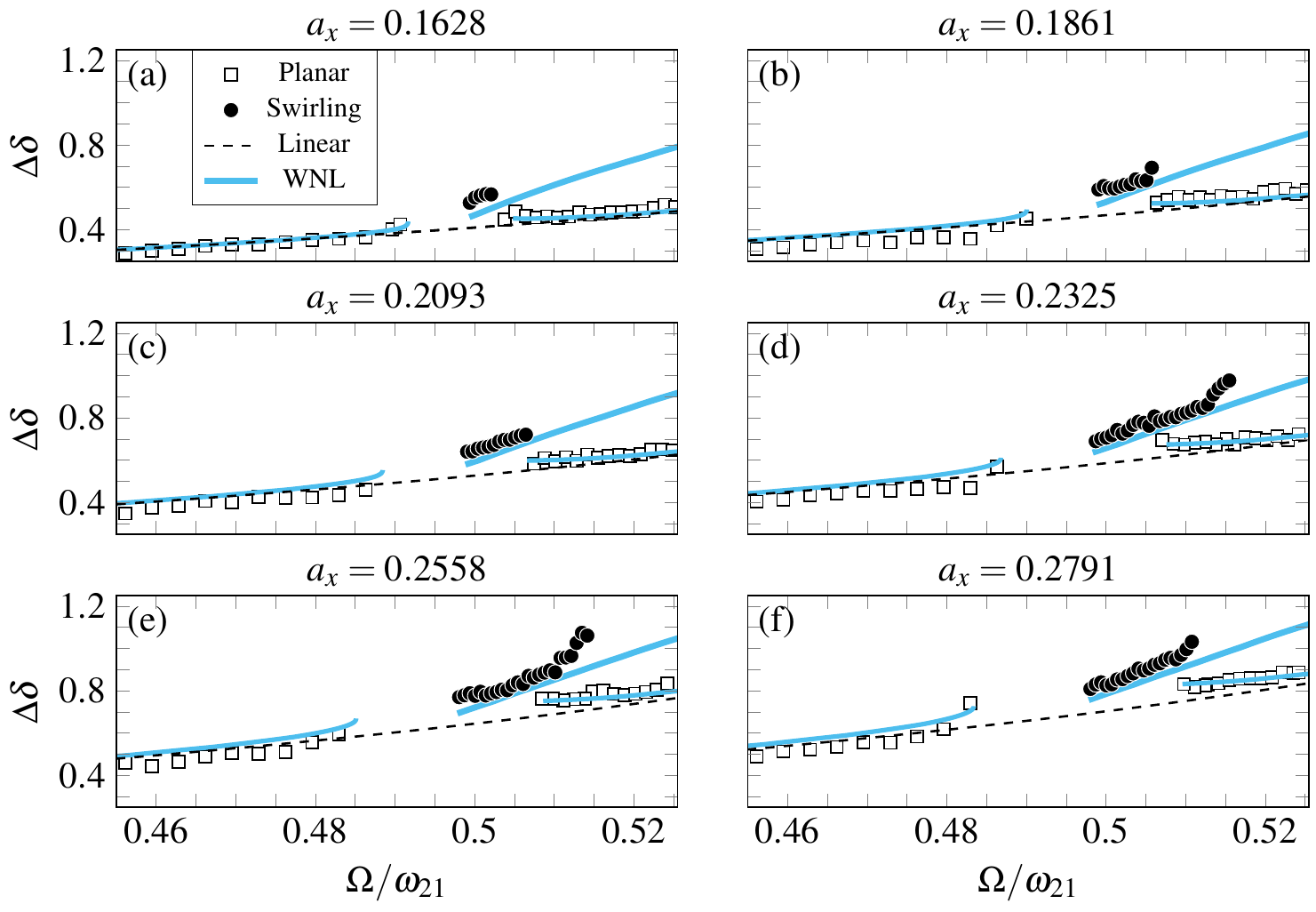}
\caption{Quantitative comparison with experimental measurements in terms of finite amplitude saturation for various non-dimensional forcing amplitudes, $a_x$. Black  dotted lines: linear potential solution according to~\eqref{eq:GenEigProb0} and~\eqref{eq:GenEigProb1}. Light blue solid lines: stable planar and swirling branches predicted by the present weakly nonlinear (WNL) model according to~\eqref{eq:DC_sol_reconst}. Markers: experimental measurements. Black empty squares correspond to planar motion whereas black filled circles refer to swirling dynamics.}
\label{fig:Fig9} 
\end{figure}

In this last section, we provide a more quantitative comparison in terms of wave amplitude saturation between the theoretical predictions according to~\eqref{eq:DC_sol_reconst} and the experimental measurements. On this point, the dimensional wave amplitude, $\Delta \bar{\delta}=\max_{\theta,t}\eta\left(r=R,\theta,t\right)-\min_{\theta,t}\eta\left(r=R,\theta,t\right)$, is experimentally measured by fixing the forcing amplitude while operating a frequency sweep in two directions. A backward sweep is used so as to follow the right lower planar branch until the sub-critical jump-up transition to swirling (P: Poincaré bifurcation) occurs ($\Omega=\Omega_{P}\left(a_x\right)$). On the other hand, an upward sweep %starting from the left lower planar branch
is performed in order to maintain a stable super-harmonic swirling response from bifurcation point $H$ ($\Omega=\Omega_{H}\left(a_x\right)$) and beyond the threshold frequency $\Omega=\Omega_{P}\left(a_x\right)$, above which the super-harmonic planar and swirling motions are both stable solutions (right region in the stability chart of figure~\ref{fig:Fig_stability_regions}).\\
\indent For each set of forcing parameters $(a_x, \Omega/\omega_{21})$, the height in pixel of the wave crest and trough on the front wall are manually extracted from the corresponding movies and compared to the height of the fluid at rest (flagged by a black mark on the container, also used as a scale) to obtain the maximal and minimal front contact line elevation. These values are converted into meters using the conversion factor provided by the black scale. The resulting difference $\Delta \bar{\delta}$ is then averaged over 3 to 5 cycles of oscillations and then normalized by the container radius $R$.\\
\indent The experimental dimensionless wave amplitude $\Delta \delta = \Delta \bar{\delta}/R$ as a function of the forcing frequency for various forcing amplitudes is displayed in figure~\ref{fig:Fig9} together with the theoretical weakly nonlinear prediction~\eqref{eq:DC_sol_reconst} (light blue solid lines) and with the linear potential solution~\eqref{eq:GenEigProb0} for comparison (black dashed line).\\
\indent The experimental data associated with the two planar branches compare generally well with the present weakly nonlinear prediction, \textcolor{black}{although the WNL theory slightly underestimates the wave amplitude in the swirling regime.} We recall from \S\ref{subsec:Sec5sub1} that, at leading order, the wave solution in these two branches is made by the superposition of two planar waves, i.e. an harmonic planar single-crest component, oscillating in space and time as $\cos{\left(\Omega t\right)}\cos{\theta}$, and a super-harmonic planar double-crest component, characterized by $\cos{\left(2\Omega t+\Phi\right)}\cos{2\theta}$, with a phase $\Phi=\pi$ in the left branch and $\Phi=0$ in the right one. The information on the phase $\Phi$ is not directly discernible from the amplitude plot of figure~\ref{fig:Fig9}, but it is contained in the snapshots sequence reported in figure~\ref{fig:fig_regular_regimes}(a) for $\Omega/\omega_{21}<0.5$ and (e) for $\Omega/\omega_{21}>0.5$. Due to the temporal periodicity of the single-crest wave, snapshots taken at $t=T/4=\pi/2\Omega$ and $t=3T/4=3\pi/2\Omega$ represent temporal nodes for the harmonic component, so that, as a first order approximation, only the double-crest component, whose azimuthal spatial structure reads $\cos{\left(\pi+\Phi\right)}\cos{2\theta}$, is instantaneously left. It is then clear that for $\Omega/\omega_{21}<0.5$ and $\Phi=\pi$, the free surface maximum is reached at the azimuthal coordinates $\theta=0$ and $\pi$, whereas the minimum is at $\theta=\pm \pi/2$ (vice versa for $\Omega/\omega_{21}>0.5$ and $\Phi=0$). This produces the concave and convex shapes in the instantaneous free surface displayed in figure~\ref{fig:fig_regular_regimes}(a) and (e), respectively.\\
%\indent Note that measurements corresponding to the irregular regime are not reported in figure~\ref{fig:Fig9} for the sake of clarity.\\
\indent Consistently with the stability chart in figure~\ref{fig:Fig_stability_regions} obtained through a backward frequency sweep, the threshold frequency $\Omega_H\left(a_x\right)$, at which the swirling branch becomes stable from lower driving frequency $\Omega$, is correctly detected. Furthermore, the upward sweep allows us to detect also the jump-down transition from the swirling to the lower right planar branch.\\
\indent The occurrence of the jump-down transition was to be expected as it is produced by dissipative mechanisms (see also \S\ref{subsec:Exp_est_bounds}), which are overlooked by the present inviscid analysis. The associated damping, which is a function of the wave amplitude and of the forcing acceleration amplitude (see \cite{raynovskyy2018damped},\cite{raynovskyy2020sloshing} and the discussion in Appendix~A of \cite{bongarzone2022amplitude}), is responsible for the modulation in the phase lag between the external driving and the wave response, which was shown by \cite{bauerlein2021phase} (for unidirectional sloshing waves in a rectangular container) to be of crucial importance for a correct prediction of the jump-down frequency.\\
\indent The damping coefficient could be tentatively fitted from experiments and phenomenologically introduced \textit{a posteriori} in amplitude equations~\eqref{eq:AmpEqDCfinal_A}-\eqref{eq:AmpEqDCfinal_B} as done in Appendix~A of \cite{bongarzone2022amplitude}. Nevertheless, the jump-down transition in the cases examined in this section (see figure~\ref{fig:Fig9}) was seen to be extremely sensitive to the frequency sweeping rate. A decrease in the frequency step increment from 5 mHz to 1 mHz (used to produced the swirling branch in figure~\ref{fig:Fig9}) was observed to give different jump-down frequencies. This is also expected as it is known from the literature that in the multi-solution range, the characteristic of the response mainly depends on the sweep rate \citep{park2011slow,bourquard2019comment,yu2020capture}. Since we did not try frequency increments smaller than 1 mHz, the jump-down frequency predictions as shown in figure~\ref{fig:Fig9} are not entirely reliable for fitting the damping at stake in the experiments.\\
\indent In spite of such limitations, the weakly nonlinear model is seen to describe fairly well the experimental swirling branch until the measured jump-down frequency. A relatively small departure of the swirling response from the theoretical prediction is typically observed at larger driving amplitude for increasing wave frequency. In agreement with previous studies \citep{dodge1965liquid,ibrahim2009liquid,bauerlein2021phase}, our experiments reveal that this is due to the progressive steepening and broadening of the wave crest and troughs, respectively, in the vicinity of the container wall. This nonlinear mechanism eventually becomes strong enough for the weakly nonlinear model to lose accurateness.   
%\cite{bongarzone2022numerical}

\section{Conclusion}\label{sec:Sec7}

In this work, the behaviour of sloshing waves in a cylindrical container submitted to longitudinal periodic forcing with driving amplitude $a_x$ and angular frequency $\Omega$ was investigated. While previous studies of this forcing condition and geometry mostly focused on the investigation of the free surface response in the vicinity of harmonic resonance, i.e. $\Omega/\omega_{1n}\approx1$, the core of the present work was dedicated to the most relevant secondary super-harmonic resonances $\Omega/ \omega_{2n} \approx 1/2$, characterized by the occurrence of a double-crest (DC) dynamics oscillating at a frequency $\omega = 2\Omega \approx \omega_{2n}$.

\indent Such a super-harmonic resonance was first experimentally observed by \cite{reclari2013hydrodynamics} and \cite{reclari2014surface} for rotary container motions, but its investigation under different forcing conditions, e.g. longitudinal forcing, seemed to be still unreported. 

\indent With the aim to take a further step in this direction, a weakly nonlinear analysis (WNL) via multiple timescale method together with a dedicated experimental compaign were implemented in order to account for the steady-state free surface dynamics, and for the symmetry-breaking due to the emergence of a double crest swirling wave in the vicinity of the super-harmonic resonance.

\indent In similar fashion to \cite{bongarzone2022amplitude}, the WNL analysis was first formalized to tackle the simpler case of harmonic resonances. The outcomes of the model were compared to previous experimental measurements and to former theoretical predictions based on the Narimanov--Moiseev multimodal sloshing theory \citep{faltinsen2016resonant,raynovskyy2020sloshing}. All together, our analysis addressing the single-crest (SC) wave dynamics was shown to be consistent with the previously reported experimental and theoretical results. In particular, the WNL model successfully captured the regime bounds between single-crest planar, swirling and irregular waves, and correctly described the close-to-resonance nonlinear behaviour, thus validating the relevance of this theoretical approach.

\indent The WNL analysis was then extended to the more complex case of the super-harmonic resonance. A dedicated lab-scale experiment was set-up to observe and characterise the super-harmonic response to longitudinal forcing. In remarkable agreement with the outcomes of the WNL model, the experimental investigation showed that the free surface dynamics in the vicinity of the super-harmonic resonance results from the superposition of a permanent, first-order forced harmonic planar single-crest wave, and of a super-harmonic double-crest wave that can exhibit either a planar, irregular or swirling dynamics, the latter being responsible for a symmetry-breaking in the system’s response through equally probable clockwise or anti-clockwise swirling waves. The bounds in the $(a_x, \Omega/\omega_{21})$ plane  between the three different regimes were experimentally retrieved and were shown to be in very good quantitative agreement with the WNL predictions, at least above a threshold forcing amplitude, below which the swirling and irregular dynamics appear to be suppressed by dissipative mechanisms, which are not accounted for by the present inviscid analysis. Finally, the predicted wave amplitude saturation, computed by reconstructing the total flow solution, was compared to the experimentally measured steady-state wave amplitude and was shown to correctly describe the stable planar and swirling branches in the neighbourhood of the super-harmonic resonance. %% to be written differently, this is almost the exact same wording as in your former paper.  

% To the knowledge of the authors, the formalization of an amplitude equation model describing the super-harmonic DC sloshing dynamics in longitudinally shaken containers and validated with a thorough dedicated experimental campaign, has not been reported in the literature yet, hence representing the most significant finding of this work. %%

The fairly good agreement between the theoretical predictions and the experimental findings validates the relevance of the WNL approach to successfully describe the sloshing wave dynamics resulting from nonlinear harmonic and super-harmonic interactions. As discussed in Appendix \ref{sec:AppC}, this analysis is not restricted to longitudinal forcing, but can be straightforwardly generalized without any further calculation to any elliptic trajectory, hence recovering the limit of rotary sloshing investigated in \cite{bongarzone2022amplitude}. In this respect, the theory of \cite{faltinsen2016resonant} for elliptical container motions interestingly predicts the occurrence of counter-rotating swirling waves, i.e. propagating in the direction opposed to that of the container motion. The qualitative analogy between the harmonic and super-harmonic system behaviour outlined in this manuscript would suggest that such counter-propagating swirling waves could also be triggered by exciting the system in the vicinity of the super-harmonic double-crest resonance, thus calling for new experimental campaigns.

\appendix

\section{Computation of the normal form coefficients}\label{sec:AppB}

\begin{table}
\centering
\begin{tabular}{c|cccccc|cccccccccccc}
$H=h/R$ & $\mu_{_{SC}}$ & & $\nu_{_{SC}}$ & & $\xi_{_{SC}}$ & & & $\mu_{_{DC}}$ & & $\nu_{_{DC}}$ & & $\xi_{_{DC}}$ & & $\zeta_{_{DC}}$ & & $\chi_{_{-}}$ & & $\chi_{_{+}}$\\ \hline
1.10 & -0.279 & & 1.414 & & -7.487 & & & 0.118 & & 9.821 & & -32.077 & & 0.104 & & 2.697 & & -3.257\\
1.20 & -0.280 & & 1.407 & & -7.914 & & & 0.108 & & 9.812 & & -32.110 & & 0.067 & & 2.692 & & -3.159\\
1.30 & -0.281 & & 1.406 & & -8.101 & & & 0.101 & & 9.813 & & -32.128 & & 0.046 & & 2.687 & & -3.089\\
1.40 & -0.282 & & 1.407 & & -8.211 & & & 0.096 & & 9.812 & & -32.138 & & 0.035 & & 2.682 & & -3.040\\
1.50 & -0.283 & & 1.409 & & -8.281 & & & 0.093 & & 9.811 & & -32.143 & & 0.029 & & 2.678 & & -3.006\\
1.60 & -0.283 & & 1.410 & & -8.328 & & & 0.091 & & 9.811 & & -32.146 & & 0.028 & & 2.675 & & -2.982\\
1.70 & -0.283 & & 1.411 & & -8.359 & & & 0.089 & & 9.810 & & -32.148 & & 0.029 & & 2.673 & & -2.965\\
1.80 & -0.284 & & 1.412 & & -8.381 & & & 0.089 & & 9.810 & & -32.149 & & 0.032 & & 2.672 & & -2.953\\
1.90 & -0.284 & & 1.412 & & -8.395 & & & 0.088 & & 9.810 & & -32.149 & & 0.035 & & 2.671 & & -2.945\\ 
2.00 & -0.284 & & 1.413 & & -8.405 & & & 0.087 & & 9.810 & & -32.150 & & 0.040 & & 2.670 & & -2.940\\ \hline
\end{tabular}
\caption{Value of the normal form coefficients appearing in~\eqref{eq:AmpEqSCfinalA}-\eqref{eq:AmpEqSCfinalB} (SC) and in~\eqref{eq:AmpEqDCfinal_A}-\eqref{eq:AmpEqDCfinal_B} (DC) computed at different fluid depths $H=h/R$ and associated with mode $\left(m,n\right)=\left(1,1\right)$. Note that in~\eqref{eq:AmpEqDCfinal_A}-\eqref{eq:AmpEqDCfinal_B}, $\chi_{_{DC}}=\chi_{_{-}}+\chi_{_{+}}$.}
\label{tab:Tab3}
\end{table}

\indent The normal form coefficients appearing in~\eqref{eq:AmpEqSCfinalA}-\eqref{eq:AmpEqSCfinalB} for the harmonic single-crest (SC) dynamics are computed as follows
\begin{subequations}
\begin{equation}
\label{eq:SC_eps3_coeff_mu}
\text{i}\,\mathcal{I}_{_{SC}}\,\mu_{_{SC}}=<\hat{\mathbf{q}}_1^{A_1 \dagger},\boldsymbol{\hat{\mathcal{F}}}_3^F>=\int_{0}^{1}\left(r/2\right)\overline{\hat{\eta}}_1^{A_1 \dagger}\, r\text{d}r,
\end{equation}
\begin{equation}
\label{eq:SC_eps3_coeff_nu}
\text{i}\,\mathcal{I}_{_{SC}}\,\nu_{_{SC}}=<\hat{\mathbf{q}}_1^{A_1 \dagger},\boldsymbol{\hat{\mathcal{F}}}_3^{|A_1|^2A_1}>=\int_{0}^{1}\left(\overline{\hat{\eta}}_1^{A_1 \dagger}\hat{\mathcal{F}}_{3_{\text{dyn}}}^{|A_1|^2A_1}+\overline{\hat{\Phi}}_1^{A_1 \dagger}\hat{\mathcal{F}}_{3_{\text{kin}}}^{|A_1|^2A_1}\right)\, r\text{d}r,
\end{equation}
\begin{equation}
\label{eq:SC_eps3_coeff_xi}
\text{i}\,\mathcal{I}_{_{SC}}\,\xi_{_{SC}}=<\hat{\mathbf{q}}_1^{A_1 \dagger},\boldsymbol{\hat{\mathcal{F}}}_3^{|B_1|^2A_1}>=\int_{0}^{1}\left(\overline{\hat{\eta}}_1^{A_1 \dagger}\hat{\mathcal{F}}_{3_{\text{dyn}}}^{|B_1|^2A_1}+\overline{\hat{\Phi}}_1^{A_1 \dagger}\hat{\mathcal{F}}_{3_{\text{kin}}}^{|B_1|^2A_1}\right)\, r\text{d}r.
\end{equation}
\end{subequations}
\noindent where $\mathcal{I}_{_{SC}}=<\hat{\mathbf{q}}_1^{A_1 \dagger},\mathsfbi{B}\hat{\mathbf{q}}_1^{A_1}>=\int_{0}^{1}\left(\overline{\hat{\eta}}_1^{A_1 \dagger}\hat{\Phi}_1^{A_1}+\overline{\hat{\Phi}}_1^{A_1 \dagger}\hat{\eta}_1^{A_1}\right)\, r\text{d}r$. Here $\left(\hat{\mathbf{q}}_1^{A_1 \dagger},\hat{\mathbf{q}}_1^{B_1 \dagger}\right)=\left(\overline{\hat{\mathbf{q}}}_1^{A_1},\overline{\hat{\mathbf{q}}}_1^{B_1}\right)$, since the inviscid problem is self--adjoint with respect to the Hermitian scalar product $<\mathbf{a},\mathbf{b}>=\int_{\Sigma}^{}\overline{\mathbf{a}}\cdot\mathbf{b}\,\text{d}V$, with $\mathbf{a}$ and $\mathbf{b}$ two generic vector (see \cite{Viola2018a} for a thorough discussion and derivation of the adjoint problem).\\ \indent Expressions~\eqref{eq:SC_eps3_coeff_mu} and~\eqref{eq:SC_eps3_coeff_nu} were already given in \cite{bongarzone2022amplitude}. The left-hand-side of those expression was typed mistakenly, as the mass matrix $\mathsfbi{B}$ should not appear in their numerators. The present version is instead written down correctly.\\
\indent For the calculation of the amplitude equation coefficients at $\epsilon^3$ order, only resonant terms matter. These terms, with their corresponding amplitudes, are proportional to $e^{\text{i}\left((\omega_{1n}t \pm \theta\right)}$ for SC waves and to $e^{\text{i}\left(\omega_{2n}t \pm 2\theta\right)}$ for DC waves. As an example, the expression of $\hat{\mathcal{F}}_{3_{kin}}^{|A|^2A}$, with $A=A_1$ for SC waves and $A=A_2$ for DC waves, is given in Appendix~D of \cite{bongarzone2022amplitude}. \indent The extraction of resonant terms was performed by using tools of symbolic calculus, e.g. the software Wolfram Mathematica.\\
\indent Analogously, the normal form coefficients appearing in~\eqref{eq:AmpEqDCfinal_A}-\eqref{eq:AmpEqDCfinal_B} for the super-harmonic double-crest (DC) dynamics are calculated as
\begin{subequations}
\begin{equation}
\label{eq:DC_eps2_coeff_mu}
\text{i}\,\mathcal{I}_{_{DC}}\,\mu_{_{DC}}=\int_{0}^{1}\left(\overline{\hat{\eta}}_1^{A_2 \dagger}\hat{\mathcal{F}}_{2_{\text{dyn}}}^{F^2}+\overline{\hat{\Phi}}_1^{A_2 \dagger}\hat{\mathcal{F}}_{2_{\text{kin}}}^{F^2}\right)\, r\text{d}r,
\end{equation}
\begin{equation}
\label{eq:DC_eps3_coeff_zeta}
\text{i}\,\mathcal{I}_{_{DC}}\,\zeta_{_{DC}}=\int_{0}^{1}\left(\overline{\hat{\eta}}_1^{A_2 \dagger}\hat{\mathcal{F}}_{3_{\text{dyn}}}^{\Lambda F^2}+\overline{\hat{\Phi}}_1^{A_2 \dagger}\hat{\mathcal{F}}_{3_{\text{kin}}}^{\Lambda F^2}\right)\, r\text{d}r,
\end{equation}
\begin{equation}
\label{eq:DC_eps3_coeff_chi}
\text{i}\,\mathcal{I}_{_{DC}}\,\chi_{_{DC}}=\int_{0}^{1}\left(\overline{\hat{\eta}}_1^{A_2 \dagger}\hat{\mathcal{F}}_{3_{\text{dyn}}}^{A_2|F|^2}+\overline{\hat{\Phi}}_1^{A_2 \dagger}\hat{\mathcal{F}}_{3_{\text{kin}}}^{A_2|F|^2}\right)\, r\text{d}r,
\end{equation}
\begin{equation}
\label{eq:DC_eps3_coeff_nu}
\text{i}\,\mathcal{I}_{_{DC}}\,\nu_{_{DC}}=\int_{0}^{1}\left(\overline{\hat{\eta}}_1^{A_2 \dagger}\hat{\mathcal{F}}_{3_{\text{dyn}}}^{|A_2|^2A_2}+\overline{\hat{\Phi}}_1^{A_2 \dagger}\hat{\mathcal{F}}_{3_{\text{kin}}}^{|A_2|^2A_2}\right)\, r\text{d}r,
\end{equation}
\begin{equation}
\label{eq:DC_eps3_coeff_nu}
\text{i}\,\mathcal{I}_{_{DC}}\,\xi_{_{DC}}=\int_{0}^{1}\left(\overline{\hat{\eta}}_1^{A_2 \dagger}\hat{\mathcal{F}}_{3_{\text{dyn}}}^{|B_2|^2A_2}+\overline{\hat{\Phi}}_1^{A_2 \dagger}\hat{\mathcal{F}}_{3_{\text{kin}}}^{|B_2|^2A_2}\right)\, r\text{d}r,
\end{equation}
\end{subequations}
\noindent with $\mathcal{I}_{_{DC}}=<\hat{\mathbf{q}}_1^{A_2 \dagger},\mathsfbi{B}\hat{\mathbf{q}}_1^{A_2}>=\int_{0}^{1}\left(\overline{\hat{\eta}}_1^{A_2 \dagger}\hat{\Phi}_1^{A_2}+\overline{\hat{\Phi}}_1^{A_2 \dagger}\hat{\eta}_1^{A_2}\right)\, r\text{d}r$. The integrals are all evaluated at the free surface $z=0$.\\
\indent We note that the value of the normal form coefficient $\chi_{_{DC}}$ contains two different contributions. Indeed, it could be conveniently rewritten as $\chi_{_{DC}}=\chi_{_{-}}+\chi_{_+}$, with the value of $\chi_{_-}$ and $\chi_{_+}$ given in table~\ref{tab:Tab3}. $\chi_{_-}$ precisely corresponds to the coefficient $\chi_{_{DC}}$ computed in \cite{bongarzone2022amplitude} and, by adopting the present formalism, e.g. for mode $A_2$ (same for mode $B_2$), it is produced by the interaction of the second order responses 
\begin{equation}
\left(1/2\right)A_2F\hat{\mathbf{q}}_2^{A_2F}e^{\text{i}\left(\left(3\omega_{2n}/2\right)t-3\theta\right)}e^{\text{i}\Lambda T_1}+\left(1/2\right)A_2\overline{F}\hat{\mathbf{q}}_2^{A_2\overline{F}}e^{\text{i}\left(\left(\omega_{2n}/2\right)t-\theta\right)}e^{-\text{i}\Lambda T_1},
\end{equation}
\noindent in equation~\eqref{eq:DC_eps2_full_sol} with the complex conjugate of the leading order particular solution characterized by $m=-1$ in~\eqref{eq:DC_sol_eps1}. On the contrary, the contribution $\chi_{_+}$ is the result of the interaction between the second order responses
\begin{equation}
\left(1/2\right)A_2F\hat{\mathbf{q}}_2^{A_2F}e^{\text{i}\left(\left(3\omega_{2n}/2\right)t-\theta\right)}e^{\text{i}\Lambda T_1}+\left(1/2\right)A_2\overline{F}\hat{\mathbf{q}}_2^{A_2\overline{F}}e^{\text{i}\left(\left(\omega_{2n}/2\right)t-3\theta\right)}e^{-\text{i}\Lambda T_1},
\end{equation}
\noindent in equation~\eqref{eq:DC_eps2_full_sol} and the complex conjugate of the leading order particular solution for $m=+1$ in~\eqref{eq:DC_sol_eps1}.\\

\section{Generalization to elliptic orbits}\label{sec:AppC}

In this appendix, we show how the analysis outlined in this manuscript for longitudinal container motions can be straightforwardly generalized to any elliptic-like shaking. For elliptical orbits in the horizontal $\left(x,y\right)$--plane, equations~\eqref{eq:EqMotWall} are modified as follows
\begin{equation}
\label{eq:EqMotWall_ELL}
\dot{\mathbf{X}}_0 =
  \begin{cases}
    \, \, \left(-a_x\Omega\sin{\left(\Omega t\right)}\cos\theta+a_y\Omega\cos{\left(\Omega t\right)}\sin\theta\right)\,\mathbf{e}_r\\
   \, \, \left(\, \, \, \, a_x\Omega\sin{\left(\Omega t\right)}\sin\theta+a_y\Omega\cos{\left(\Omega t\right)}\sin\theta\right)\,\mathbf{e}_{\theta}
  \end{cases},
\end{equation}
with $a_x$ and $a_y$ the non-dimensional major- and minor-axis forcing amplitude components, respectively, and $\Omega$ the non-dimensional driving angular frequency. Under these forcing conditions, the unsteady and forced Bernoulli's equation at $z=\eta$ reads
\begin{equation}
\label{eq:GovEq_Dyn_APP}
\frac{\partial\Phi}{\partial t}+\frac{1}{2}\nabla\Phi\cdot\nabla\Phi+\eta=r\left(f_x\cos{\left(\Omega t\right)}\cos{\theta}+f_y\sin{\left(\Omega t\right)}\sin{\theta}\right),
\end{equation}
where $f_x=a_x\Omega^2$ and $f_y=a_y\Omega^2$. By introducing the aspect ratio $\alpha=a_y/a_x=f_y/f_x$, so that $f_x=f$ and $f_y=\alpha f$, equation~\eqref{eq:GovEq_Dyn_APP} can be conveniently rewritten as
\begin{equation}
\label{eq:GovEq_Dyn_new}
\frac{\partial\Phi}{\partial t}+\frac{1}{2}\nabla\Phi\cdot\nabla\Phi+\eta=r\frac{f}{2}\left(\left(\frac{1+\alpha}{2}\right)e^{\text{i}\left(\Omega t-\theta\right)}+\left(\frac{1-\alpha}{2}\right)e^{\text{i}\left(\Omega t+\theta\right)}\right)+c.c.\,.
\end{equation}
A value $0<\alpha<1$ implies elliptic orbits, whereas the two limit cases with $\alpha=0$ ($a_x\ne0$, $a_y=0$) and $\alpha=1$ ($a_x=a_y\ne0$) correspond, respectively, to longitudinal, as in the present work, and rotary \citep{bongarzone2022amplitude}, shaking conditions. For convenience of notation, we also introduce the auxiliary variables
\begin{equation}
\label{eq:alphapm}
\alpha_{_-}=\frac{1+\alpha}{2},\ \ \ \ \ \ \ \ \alpha_{_+}=\frac{1-\alpha}{2},
\end{equation}
with $1/2\le\alpha_{_-}\le1$ and $0\le\alpha_{_+}\le1/2$. By accounting for the two auxiliary aspect ratios, $\alpha_{_-}$ and $\alpha_{_+}$ in the expression of the forcing term, the whole derivation can be repeated, hence leading, without any further computation, to the following system of amplitude equations for harmonic single-crest (SC) waves
\begin{subequations}
\begin{equation}
\label{eq:AmpEqSCfinalA_elliptic}
\frac{dA}{dt}=-\text{i}\lambda A + \text{i}\,\mu_{_{SC}} \alpha_{_-}f + \text{i}\,\nu_{_{SC}} |A|^2A +\text{i}\,\xi_{_{SC}}|B|^2A,
\end{equation}
\begin{equation}
\label{eq:AmpEqSCfinalB_elliptic}
\frac{dB}{dt}=-\text{i}\lambda B + \text{i}\,\mu_{_{SC}} \alpha_{_+}f + \text{i}\,\nu_{_{SC}} |B|^2B + \text{i}\,\xi_{_{SC}}|A|^2B.
\end{equation}
\end{subequations}
\noindent and for super-harmonic double-crest (DC) waves
\begin{subequations}
\begin{eqnarray}
\label{eq:AmpEqDCfinal_A_elliptic}
\frac{dA}{dt}=-\text{i}\,\left(2\lambda-\left(\alpha_{_-}^2\chi_{_-}+\alpha_{_+}^2\chi_{_+}\right) f^2\right) A + \text{i}\,\left(\zeta_{_{DC}}\lambda + \mu_{_{DC}}\right) \alpha_{_-}^2 f^2 \notag \\ +\text{i}\,\nu_{_{DC}} |A|^2A+\text{i}\,\xi_{_{DC}}|B|^2A,
\end{eqnarray}
\begin{eqnarray}
\label{eq:AmpEqDCfinal_B_elliptic}
\frac{dB}{dt}=-\text{i}\,\left(2\lambda-\left(\alpha_{_+}^2\chi_{_+}+\alpha_{_-}^2\chi_{_-}\right) f^2\right) B + \text{i}\,\left(\zeta_{_{DC}}\lambda + \mu_{_{DC}}\right) \alpha_{_+}^2 f^2 \notag \\ + \text{i}\,\nu_{_{DC}} |B|^2B+\text{i}\,\xi_{_{DC}}|A|^2B,
\end{eqnarray}
\end{subequations}
\noindent with the values of the normal form coefficients still given in table~\ref{tab:Tab3}.\\
\indent We note that in the limit of $\alpha=0$ (longitudinal), $\alpha_{_-}=\alpha_{_+}=1/2$ and equations~\eqref{eq:AmpEqSCfinalA}-\eqref{eq:AmpEqSCfinalB} and~\eqref{eq:AmpEqDCfinal_A}-\eqref{eq:AmpEqDCfinal_B} are retrieved. On the contrary, in the limit of $\alpha=1$ (rotary), one has $\alpha_{_-}=1$ and $\alpha_{_+}=0$, so that equations~\eqref{eq:AmpEqSCfinalA_elliptic} and~\eqref{eq:AmpEqDCfinal_A_elliptic} corresponds to equations~(4.6) and~(4.22) of \cite{bongarzone2022amplitude}, with $A\ne0$ and $B=0$ the only possible stable stationary solution for~\eqref{eq:AmpEqDCfinal_A_elliptic} and~\eqref{eq:AmpEqDCfinal_B_elliptic}.

\section{Estimation of the duration of the transient regime}\label{sec:AppD}

\begin{figure}
    \centering
    \includegraphics[width=0.8\textwidth]{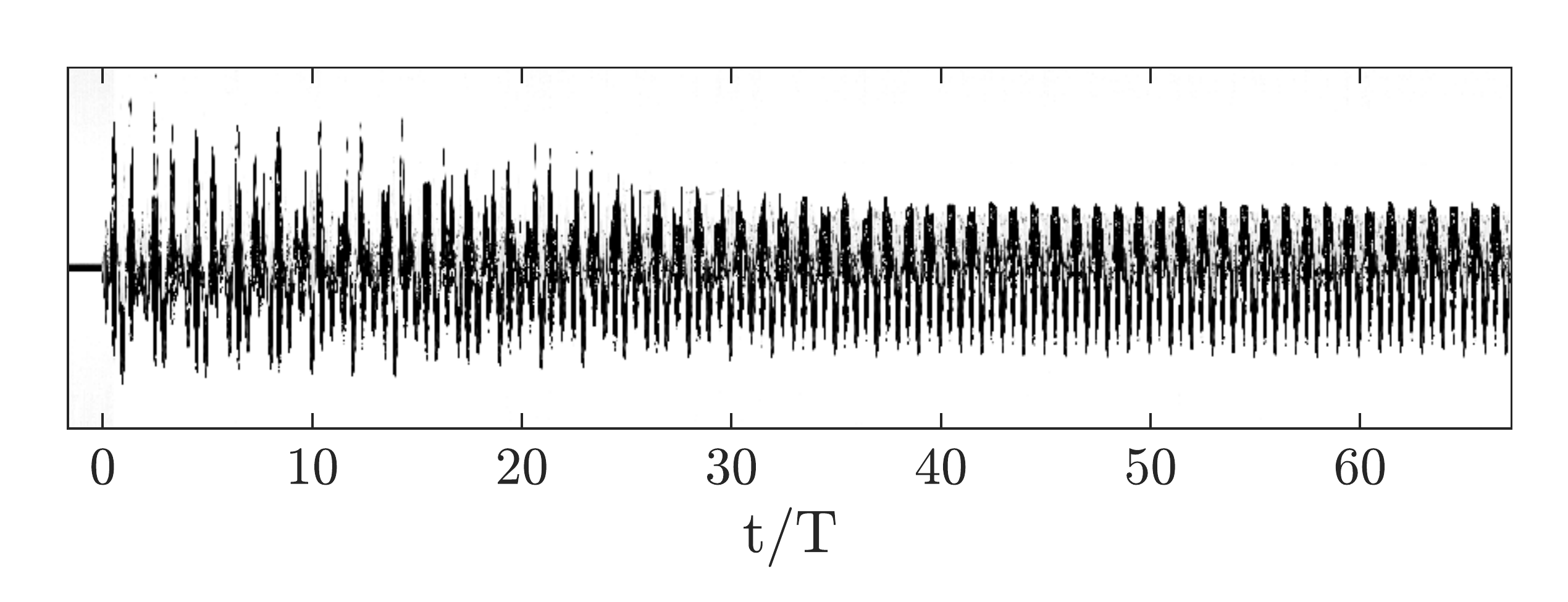}
    \caption{Intensity profile along the middle axis of the container as a function of time. The free surface, initially at rest ($t<0$) is submitted to forced harmonic oscillations from $t=0$.}
    \label{fig:regime_transitoire}
\end{figure}

In this study, we only consider the permanent response of the free surface to forced oscillations. To ensure we discard the transient regime in our analysis of the free surface dynamics, we first obtained an estimation of the transient time by recording for various forcing amplitudes $\bar{a}_x$ and angular frequencies \textcolor{black}{$\bar{\Omega}$}, the full dynamics of the free-surface, initially at rest and then put into oscillations. The temporal evolution of the intensity profile along the middle axis of the container extracted from our movies, is a direct signature of the variation in time of the sloshing wave amplitude, and reveals that for all ($\bar{a}_x$, $\bar{\Omega}$) set of parameters investigated, the free-surface dynamics can be safely considered as having reached a steady-state after typically 50 cycles of oscillations, see figure \ref{fig:regime_transitoire}. 

%\section{\textcolor{black}{Damping estimation from experiments}}\label{sec:AppD}

%%%%%%%%%%%%%%%%%%%%%%%%%%%%%%%%%%%
%%%%%%%%%%%%%%%%%%%%%%%%%%%%%%%%%%%
%%%%%%%%%%%%%%%%%%%%%%%%%%%%%%%%%%%

\subsubsection*{\textbf{\textup{Supplementary Material and Movies}}}
\indent \, Supplementary movies show the evolution of the free surface dynamics experimentally observed at increasing forcing frequency and for a fixed forcing amplitude $\bar{a}_x/R=20\,\text{mm}$, which corresponds to a non-dimensional value $a_x=\bar{a}_x/R=0.2325$. Supplementary movies are available at (LINK).
\subsubsection*{\textbf{\textup{Funding}}}
\indent \, We acknowledge the Swiss National Science Foundation under grants 178971 and 200341.
\subsubsection*{\textbf{\textup{Declaration of Interests}}}
\indent \, The authors report no conflict of interest.
\subsubsection*{\textbf{\textup{Author Contributions}}}
\indent \, A. M., F. G. and A. B. created the research plan. A.B. formulated analytical and numerical models. A.M. and A.B. led model solutions. A.M. designed and performed all experiments. A.M., F.G. and A.B. wrote the manuscript.
%\subsubsection*{\textbf{\textup{Data Availability Statement}}}
%\indent Raw data are available from the corresponding author upon request.

%%%%%%%%%%%%%%%%%%%%%%%%%%%%%%%%%%
%%%%%%%%%%% BIBLIOGRAPHY %%%%%%%%%%%%%%
%%%%%%%%%%%%%%%%%%%%%%%%%%%%%%%%%%

\bibliographystyle{jfm}
 %Note the spaces between the initials
\bibliography{Bibliography}

\end{document}